\documentclass[12pt]{article}
\usepackage[slantedGreek]{mathptmx}
\usepackage{fullpage}
\usepackage{graphicx}        % standard LaTeX graphics tool 
\usepackage{amssymb, amsmath,bm}
\usepackage{stmaryrd}
\usepackage{enumitem}
\usepackage{comment}
\usepackage{caption, rotating}
\usepackage[small]{subfigure}
\usepackage[small]{subfigure}
\usepackage{amsfonts}
\usepackage{pdflscape}
\usepackage{xcolor}
\renewcommand{\d}{\text{d}}

\numberwithin{equation}{section}
\usepackage[numbers,square,comma,sort&compress]{natbib}
\graphicspath{{./figures/}} 

\title{Analytical shear-band process zone model incorporating nonlinear viscous effects and initial defects} 

\author{J.D. Clayton$^{1}$\footnote{Email: john.d.clayton1.civ@army.mil} \\ \\
$^1$Terminal Effects Division, Army Research Directorate
\\
DEVCOM ARL, Aberdeen, MD 21005-5066, USA 
}

\date{}
\begin{document}
\graphicspath{{figures/}}
\maketitle

\begin{abstract}
Experimental, theoretical, and numerical studies of adiabatic shear in ductile
metals suggest initial defects such as pores
or material imperfections increase shear-band susceptibility. Conversely, viscous effects manifesting
macroscopically as strain-rate sensitivity inhibit localization.
The analytical shear-band process zone model due to D.E.~Grady, in turn based on a rigid-plastic solution
for stress release by N.F.~Mott, is advanced to account for these phenomena.
The material contains an average defect measure (e.g., porosity) and a concentrated defect measure
 at a spatial location where shear banding is most likely to initiate after an
instability threshold is attained.
Shearing resistance and certain physical properties are reduced commensurately with
local defect concentration. Non-Newtonian viscosity increases dissipative resistance. 
Viscous dissipation, if strong enough, is shown to prevent an infinitesimal-width shear band even
in a non-conductor.
Here, a pseudo-quadratic viscosity widens the band similarly
to heat conduction, and akin to quadratic shock viscosity often used to resolve widths
of planar shock waves. 
The model captures simulation data showing reduced localization strain and 
 shear band width with increasing maximum initial pore size in additively manufactured titanium and HY-100 steel. 
Predictions for shear band width, local strain, and temperature are more
accurate versus data on steel than prior analytical modeling.
A quantitative framework is established by which processing defects
can be related to shear-banding characteristics.
 \end{abstract}
% \noindent \textbf{Key words}: shear localization; plasticity; porosity; steel; titanium; additive manufacturing\\
\noindent \textbf{Key words}: continuum physics; materials science; adiabatic shear; plasticity; porosity; metals \\% ; additive manufacturing\\

\noindent

 %\tableofcontents

\section{Introduction}
Adiabatic shear localization is an important phenomenon in
metals deformed at high rates, whereby heat conduction is limited by short time scales.
At some threshold strain, softening by thermal, geometric, and/or
structural transformation mechanisms overtakes strain- and strain-rate hardening,
leading to instability. Localization into one or more thin bands of rapidly deforming
material of large strain and high temperature relative to surrounding material
 occurs at some time after the instability threshold.
Shear stress drops commensurately as the material unloads, and ductile fracture
often, but not always, occurs in later stages of the process.
See Refs.~\cite{wright2002,yan2021} for background.

In a torsion experiment, the macroscopic strain when this rapid stress drop initiates is
typically termed the ``critical strain'' (e.g., \cite{molinari1987}).
Strain-rate sensitivity tends to reduce shear band susceptibility. Thermal diffusivity
increases critical strain, especially at low applied strain rates \cite{wright2002}, and provides a regularization
length lending the shear band a finite width or thickness.
Inertia increases critical strain at high rates \cite{wright1996}, with momentum diffusion thought to
prevent the band from attaining an infinite width \cite{grady1992}.
While certain physics, including (an approximate lower bound on) critical strain, can be modeled
under adiabatic and quasi-static approximations \cite{molinari1987,molinari1988,wright2002,claytonJMPS2024},
shear band width and internal strain cannot be predicted under such approximations in the general case. 
In the absence of an intrinsic length scale supplied by the constitutive model (e.g., conduction
or a regularized damage-fracture model \cite{arriaga2017,claytonarx2025}), the width of the band is
dictated by boundary conditions and any initial perturbations in geometry or material properties.
While strain-rate sensitivity can provide an implicit length scale instilling unique (numerical) solutions \cite{needleman1988}, the band size is controlled by the geometry and initial conditions in the absence of intrinsic regularization.
Since heat conduction is paramount with regard to shear band width in typical metals, with none being perfect insulators, labeling such localization bands as ``adiabatic'' is an obvious misnomer \cite{wright1992,wright2002}.

In the oft-analyzed setting of simple shear of a thermoviscoplastic solid with uniform, thermally insulated far-field boundary conditions \cite{shawki1988,molinari1987,cherukuri1995,wright2002}, the homogeneous solution is the unique solution,
and localization does not occur. A local perturbation such as a geometric defect, strength heterogeneity, or temperature spike 
is needed to physically induce a spatially non-homogeneous solution to the governing equations.
Localized solutions can also be instilled, non-physically, in broader computer simulations of dynamic deformation of this class of materials by minute numerical perturbations, for example floating-point error.
Depending on the loading conditions, regions of high shear strain initiate at multiple locations if defect distributions
exist \cite{zhou2006,kubair2015,vishnu2022,vishnu2022b}. However, in typical dynamic experiments on
thin-walled tubes in a torsional split-Hopkinson (pressure) bar (i.e., SHPB or Kolsky bar) \cite{cho1990,duffy1992,fellows2001,xu2008},
and some numerical simulations of the process \cite{vishnu2022},
failure is usually dictated by behavior of one dominant shear band that grows most rapidly.
In this setting, growth of narrowly spaced, nascent bands is arrested by stress release from dominant band(s) \cite{grady1987}.

Additive manufacturing (AM) with metallic materials often produces structures with non-negligible porosity or void space \cite{sloti2014,ning2020,marvi2021}.
As might be expected, pores act as preferential sites for localized deformation, including shear banding. Mechanisms include enhanced local deformation leading to thermal softening, inter-void linking,
and ductile fracture \cite{dodd1983,batra1994,nahshon2008}.
In SHPB torsion, porous material tends to show a reduced critical strain for localization relative to fully dense material,
as evidenced by experiment \cite{silva1997} and simulation \cite{vishnu2022}.
In the latter, dominant shear bands tended to nucleate where void diameters and local void volume fractions were largest.
Although increasing the average porosity also reduced critical strain, the effect of local void concentration had a greater effect;
furthermore, larger pores are naturally more probable as the average porosity increases. Similar trends were observed in ring-compression simulations \cite{vishnu2022b}.
In such simulations \cite{vishnu2022,vishnu2022b}, as well as prior analytical studies \cite{fressengeas1987,molinari1987,molinari1988,wright2002,claytonJMPS2024}, localization was impeded by viscoplastic effects
manifested through strain-rate sensitivity.

The objective of the present study is derivation and implementation of an analytical framework for
physical characteristics of shear bands in viscous metals with initial defects (e.g., porosity from AM).
Closed-form expressions for critical strain or localization time are sought, along with expressions for
the width of the band and its average internal strain and temperature, all of which can be compared versus experimental
data (e.g., \cite{giovanola1988,duffy1992}).
While contemporary numerical simulations allow for closer representation of complex constitutive behavior
and microstructures, these are time-intensive, require software expertise, and must generally be repeated for each new material system. Relatively simple analytical expressions, if accurate, facilitate an immediate physical understanding
and suggest design guidelines. For example, if the most important material properties and parameters can be identified
in an analytical framework, then particular materials and microstructures can be targeted or optimized for a given application.
In the present context, one may seek a material to minimize or maximize the critical localization strain for a given applied strain rate and constraints on average flow stress, mass density, etc. In particular, for porous metals as produced by AM, a mathematical framework is sought by which effects of local defect concentration affect adiabatic shear.
Of course, some physical rigor must be sacrificed to produce a tractable problem yielding expressions simple enough to be evaluated without numerical iteration.

The 1-D simple shear problem is analyzed herein. As reviewed elsewhere \cite{wright2002,yan2021}, this problem has been
the subject of many prior analytical studies, though none appear to address effects of heterogeneous porosity in the manner
postulated here. The goal is \textit{not} to develop a new (3-D) constitutive model for porous ductile metals as in Refs.~\cite{gurson1977,batra1994,nahshon2008}, but rather to describe the
1-D shear-band problem using basic constitutive assumptions specialized to simple shear.
Several well-known analytical expressions for shear band width make assumptions on steadiness of the late-stage
flow process in the band \cite{dodd1989,wright1992,dinzart1998}. The steady flow assumption has been
questioned by finite-difference computations \cite{cherukuri1995}, and while these solutions all include thermal diffusivity,
and those of Refs.~\cite{wright1992,dinzart1998} include rate sensitivity, they all omit momentum diffusion.

In contrast, the analytical shear-band process zone model of Grady, most thoroughly derived in Ref.~\cite{grady1992},
accounts for dynamic stress release as well as heat conduction. The standard form of the model 
\cite{grady1991,grady1994} omits
rate dependence, but a linear viscous stress (i.e., Newtonian viscosity) was investigated briefly in Ref.~\cite{grady1992}.
Origins of the model stem from seminal work of Mott \cite{mott1947}, namely an analytical solution for
tensile expansion and release of a fragmenting cylindrical ring, later enhanced by Kipp and Grady \cite{kipp1985}.
Grady and Kipp applied such concepts to predict spacing of shear bands witnessed in shock-wave compression \cite{grady1985,grady1987}. This led to the more formal treatment of the shear-band evolution process
and shear-band toughness by Grady \cite{grady1991,grady1992,grady1994}.
A recent theoretical study \cite{sheng2024} extended the rate-independent models of Grady and Kipp to include an intermediate
viscoplastic unloading zone and rate dependent response of the far-field plastic material.
By including these extra features, closer agreement with certain data was reported \cite{sheng2024}; for
example, the original model \cite{grady1992} tended to predict narrower bands of higher strain
than witnessed experimentally \cite{giovanola1988}, whereas viscoplasticity widens the band and reduces its maximum strain. However,  additions in Ref.~\cite{sheng2024} 
complicate the analysis and final expressions, and values to be used for some of the constants and
factors in the solution are unclear or are not reported. As discussed later, the model
of Ref.~\cite{sheng2024} would also seem to severely overestimate temperature rise in the band.

In the current work, Grady's theory \cite{grady1992} is newly augmented with a non-Newtonian viscosity for the shear-slip zone to
account for rate dependence and provide a more diffuse shear band as seen experimentally.
The particular form of viscosity is quadratic, but the viscosity coefficient is normalized by the ambient background strain rate
such that a linear form is recovered when the shear band strain rate matches that of the background.
The value of the viscosity coefficient for slip discontinuity correlates with strain-rate sensitivity
of the continuum.
Noting that strain-rate sensitivity exponents are much smaller than unity, the model does \textit{not} surmise a linear or quadratic viscosity for the background continuum, but only for the local slip discontinuity.
The proposed increasing viscous stress with increasing slip velocity could be justified, conceptually, by
the increase in strain-rate sensitivity with rate witnessed in experiments on similar metals to those studied here \cite{chiou2007,sadjadpour2015,claytonJMPS2024}.
The derived solution for shear band characteristics contains a dependence on viscosity coefficient that is nearly identical to dependence on thermal diffusivity. Unlike the linear viscosity introduced briefly by Grady \cite{grady1992}, the
nonlinear viscosity enables a finite shear band width even in the absence of conduction, and it further allows for convenient closed-form expressions precluded by a linear viscosity or a viscoplastic continuum \cite{sheng2024}. 
The quadratic viscosity is similar to the quadratic shock viscosity used to lend a finite width to shock waves when
viscous dissipation is otherwise absent \cite{benson2007,mattsson2015}.
Although often termed ``artificial'', such shock viscosity can be justified to represent real physical behavior (e.g., in gas dynamics \cite{mattsson2015}). Analogously, the viscosity assigned herein can be used to represent the true observed width of a shear band,
even in cases where the underlying material is fully adiabatic as in numerical simulations \cite{vishnu2022,vishnu2022b}.
Viscosity thereby lends regularization from dissipative microscopic processes not captured by heat conduction. 

A second new component of the proposed model accounts for initial defect concentrations.  The dominant shear band is
assumed to initiate at the instability threshold \cite{anand1987,wright1996,wright2002} or at some finite time thereafter, at the spatial location where the defect density is highest relative to the average defect content (e.g., average porosity) of the background material, as corroborated by
numerical studies \cite{batra1994,vishnu2022,vishnu2022b}. Weakening of the material in the slip band is accelerated by
the defect concentration, manifesting by a linear damage-type model whose single parameter controls the rate of degradation.
For a porous material, such degradation manifests from increased geometrical softening and possible void linkage and fracture.
Viscous forces and thermal diffusivity are also reduced by defects. The model preserves the simple closed-form
expressions for band characteristics derived by Grady \cite{grady1992}, with complexity encompassed by
relating the damage constitutive parameter to initial microstructure or geometry. Herein, both the viscous and defect-related
parameters are quantified using numerical data from simulations of dynamic torsion SHPB tests on titanium and steel with discrete voids \cite{vishnu2022}. Although the data present significant scatter, the model correctly captures the trend of reduced critical strain, and a narrower band, with increasing local defect concentration.
% A thinner band of lower final temperature is predicted as defect concentration increases, implying a more brittle response with increasing local porosity. 
A very recent generalization \cite{chen2025} of Grady's model allows
strength degradation from microstructure processes such as dynamic recrystallization (DRX; cf.~\cite{rittel2008}); effects of thermal- and microstructure-softening on dissipated energy are quantified via temperature measurements.

Sections 2, 3, and 4 report the transient boundary value problem, governing equations, and their solution. Results follow in Section 5, including comparison with numerical and experimental data. Following the conclusions in Section 6, Appendix A discusses instability criteria used to specify initial conditions at nucleation, and Appendix B explains defect concentration measures.

%\pagebreak
\section{Rigid-plastic analysis}
\begin{figure}%[ht!]
\begin{center}
\includegraphics[width=0.8\textwidth]{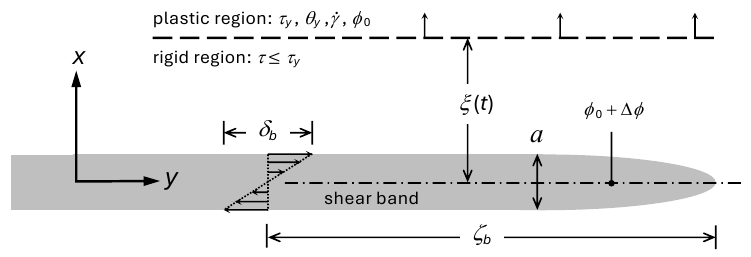} 
 % \vspace{-0.5cm}
\caption{Boundary value problem for shear localization pioneered by Grady \cite{grady1991,grady1992}.
A shear band evolves in a background material of dense flow strength $\tau_y$,
temperature $\theta_y$, strain rate $\dot{\gamma}$, and porosity $\phi_0$ newly
introduced here. Shear band of width $a$ initiates where local defect concentration is larger by $\Delta \phi$.
Shear displacement in fully formed band of process zone length $\zeta_b$
is $\delta_b = 2 \psi_c$; excess shear strain in band is $\gamma_b = \delta_b / a$.
Interfacial distance $\xi$ demarcating unloaded rigid domain
increases with time while stress relaxes in band. At the rigid-plastic interface, $\tau_0 = \tau_y = \tau / (1- \phi_0)$.
Shear band energy is $\Gamma = \tau_y \psi_c / 2$.}
\label{fig1}       
\end{center}
\end{figure}

The problem set-up is similar to that of Refs.~\cite{grady1991,grady1992}, two major exceptions being (i) the
solid may contain pores or other defects, and (ii) rate sensitivity is included.
The material, presumably a ductile polycrystalline metal, is isotropic, incompressible, and rigid-viscoplastic (i.e., negligible elastic shear strain, cf.~\cite{wright2002,molinari1987,claytonJMPS2024}).
A planar shear band forms in an effectively infinite, 2-D medium, as shown in Fig.~\ref{fig1}.
Spatial coordinates are $(x,y)$, with $|x|$ denoting distance from the center of the band.
The band can either nucleate homogeneously across the entire $y$-domain or nucleate at a specific point, say $y_0$,
and then propagate in the $+y$-direction. Thickness $a$ of the band is assumed constant \cite{grady1991,grady1992}, but
the mean shear strain $\gamma_b = \delta_b/a$ in the band increases from the start time for nucleation, labeled as the datum $t=t_0$,
to a critical time $t_c$ when the band ceases to evolve and the analysis terminates \cite{grady1992,sheng2024}.
At $t = t_c$, stress supported by the band goes to zero, and either the material melts or fractures immediately thereafter,
or the experimental apparatus such as a torsional SHPB is assumed to instantly unload without further applied strain.
Detailed modeling of the dynamics of these terminal post-failure and elastic unloading processes, for $t > t_c$, are beyond the scope 
of the current treatment.

If the band is propagating, the length of the process zone wherein $\gamma_b$ steadily increases is $\zeta_b$.
As in Refs.~\cite{grady1985,grady1987,grady1991,grady1992}, the behavior of the band itself is
collapsed to that of a singular surface at $x = 0$. The magnitude of shear-slip supported by the upper and lower halves of the band is $\psi (t)$ at arbitrary time $t$.  In the fully formed band, the shear-slip displacement is $\psi(t_c) = \psi_c = \delta_b / 2$ for $x > 0$ and $\psi_c = - \delta_b/2$ for $x <0$, with a jump of magnitude $\delta_b$ as shown
in Fig.~\ref{fig1}.  Material everywhere is either in a state of simple shear,
a rigid-body state, or at a singular boundary surface demarcating such states.
With increasing $t$, shear stress supported by the band decreases, and a planar relaxation shear wave propagates symmetrically
in the $\pm x$ directions. From symmetry, analysis of the problem can be limited to the domain $x \geq 0$.
Assuming $\zeta_b \gg \delta_b \gg a$, verified a posteriori, $y$ dependence is temporarily ignored and the problem is analyzed in $(x,t)$ space \cite{grady1991,grady1992}.

Material far away from the band is in a state of simple shear at constant shear strain rate $\dot{\gamma} > 0$, constant temperature $\theta_y$, constant porosity $\phi_0$, and constant Cauchy shear stress $\tau_y$ per unit area of the dense matrix. More realistically, all of these quantities would evolve during the shear-band evolution process, but
as in Refs.~\cite{grady1991,grady1992,sheng2024}, the far-field state is assumed stationary to render the analysis tractable, as the local state of the material
within and close to the band is assumed to change much more drastically than that of the background material. If the initial defect distribution is non-uniform, the shear band nucleates where the local defect concentration increases by an amount $\Delta \phi$. Variations in $\phi(x)$ for $2 |x| > a $ are ignored: material is viewed as homogeneous away from the band. When the band is idealized as a singular surface of shear displacement, then all effects of $\Delta \phi$ are
concentrated at $x = 0$ with $\phi = \phi_0$ for $|x| > 0$. Away from singular surfaces, continuum laws for conservation of mass, momentum, and energy are, respectively (e.g., \cite{claytonNCM2011}),
\begin{equation}
\label{eq:contlaws}
\rho = (1 - \phi_0) \rho_0, \qquad \frac{\partial \tau}{\partial x} = \rho \dot{ \upsilon}, \qquad \rho c \dot{\theta} = \beta \tau \frac{\partial \upsilon}{\partial x} + k \frac{\partial^2 \theta}{\partial x^2}.  
\end{equation}
Ambient mass density is $\rho$; density of material without voids is $\rho_0$.
Particle velocity in the $y$-direction is $\upsilon(x,t)$, and $\tau$ is the $xy$-component of Cauchy stress.
Dots denote partial time derivatives at fixed $x$, noting Eulerian and Lagrangian $x$-coordinates coincide for this problem.
A constant hydrostatic pressure could be applied, but this would not affect the analysis since the solid and voids are incompressible and
constitutive response functions are simply assumed independent of pressure.
Specific heat per unit mass is $c = \rm{const} > 0$, any distinction between constant-pressure and constant-volume irrelevant due to incompressibility. 
The Taylor-Quinney factor $\beta > 0$ denoting instantaneous fraction of plastic work converted to heat is assumed constant as is typical \cite{molinari1987,sheng2024,chen2025}, though experiment, theory, and simulation support its variability for many metals \cite{wright2002,claytonJMPS2005,rittel2006b,longere2018,kunda2024}. 

Away from the band, stress \cite{gurson1977,becker1988,claytonJDBM2021} and conductivity vary constitutively with porosity:
\begin{equation}
\label{eq:taukpores}
k = (1-\phi_0) k_0 = \rho c  \chi_0, \qquad \tau(x,t) = (1-\phi_0) \tau_0(x,t).
\end{equation}
Stress supported by the dense matrix material in the absence of voids is $\tau_0$, and $k_0$ likewise is conductivity of the dense solid.
Thermal diffusivity $\chi_0 = k/ (\rho c) = k_0 / (\rho_0 c)$ is independent of $\phi_0$.
The second and third of \eqref{eq:contlaws} become
\begin{equation}
\label{eq:contlawsred}
{\partial \tau_0}/{\partial x} = \rho_0 \dot{ \upsilon}, \qquad \rho_0 c \dot{\theta} = \beta \tau_0 ({\partial \upsilon}/{\partial x}) + k_0 ({\partial^2 \theta}/{\partial x^2}).  
\end{equation}

In the plastic region of Fig.~\ref{fig1}, $\tau_0 = \tau_y = \rm{const}$,
$\dot{\upsilon} = 0$, $\partial \upsilon / \partial x = \dot{\gamma} = \rm{const}$, and $\theta = \theta_y = \rm{const}$
during the rapid evolution of the shear band.
Strain hardening, thermal softening, and adiabatic temperature rise in the last of \eqref{eq:contlawsred} are 
omitted in the far-field region \cite{grady1992,sheng2024}, but only in the very brief time domain of localization $t \in  (t_0,t_c]$. These physics are included in the analysis for $t \leq t_0$.
In the singular description of the shear plane, stress degrades linearly with shear displacement $\psi(t)$ \cite{grady1991,grady1992} in an assumed form of the solution; the constitutive model for $\tau$ follows in \S3. Along $x = 0$,
\begin{equation}
\label{eq:taudeg}
\tau (\psi(t)) = (1-\phi_0) \tau_0 (\psi(t)) = (1-\phi_0) \tau_y (1 - \psi(t)/ \psi_c) \quad
\Rightarrow \quad \tau_0 = \tau_y (1-\psi/\psi_c),
\end{equation}
noting $\psi \in [0,\psi_c]$. Recalling $\dot{\gamma} = \rm{const} > 0$, the velocity field in the two regions of Fig.~\ref{fig1} obeys
\begin{equation}
\label{eq:velfield}
\upsilon(x) = \begin{cases} 
& \dot{\gamma} \xi, \qquad x \in [0,\xi),  \\
& \dot{\gamma} x, \qquad x \in [\xi,\infty). 
\end{cases}
\end{equation}
Linear momentum in the $y$ direction per unit area of the $x = 0$ plane of the entire body is $P$:
\begin{equation}
\label{eq:momP}
P(t) = \int_0^\infty \rho \upsilon \, \d x = (1-\phi_0)\rho_0 \dot{\gamma} \xi^2 + (1-\phi_0)\rho_0 \dot{\gamma} \int_\xi^\infty x \, \d x .
\end{equation}
In the singular-surface description of the band, $\Delta \phi$ is excluded from the integral.
Net force in the $y$ direction, per unit area, supported
by any finite region $\{B_0: 0^+ \leq x \leq R \} $, where $R \in [\xi,\infty)$, is $F(t) = \tau_y (1-\phi_0) - \tau(t)$.
A global momentum balance for $B_0$ is, dividing by $1-\phi_0$
and using \eqref{eq:taudeg},
\begin{equation}
\label{eq:globmom}
F = \frac{ \d P}{ \d t } \quad \Rightarrow \quad  \tau_y - \tau_0(t) =
\tau_y \frac{\psi(t)}{\psi_c} =
\rho_0 \dot{\gamma} \xi (t) \frac {\d \xi(t)}{ \d t}
; \qquad \frac{\d \psi(t)}{ \d t} = \dot{\gamma} \xi(t).
\end{equation}
Velocity of the slipped plane $\d \psi / \d t$ must match that of the rigid region.
For convenience, the initial datum is taken as $t_0 = 0$; thus $t$ is negative up to the start of localization.
The two differential equations in \eqref{eq:globmom} with initial conditions $\psi(0) = \xi(0) = 0$
have the immediate solution \cite{grady1987}
\begin{align}
\label{eq:rigidsol}
\psi(t) = \frac{\tau_y \dot{\gamma}}{18 \rho_0 \psi_c} t^3, \qquad \xi(t) = \frac{\tau_y}{6 \rho_0 \psi_c} t^2; \qquad
\psi^2_c = \frac{\tau_y \dot{\gamma}}{18 \rho_0 } t_c^3.
\end{align}
Note the slipped distance at stress collapse, $\psi_c = \psi(t_c)$, is an outcome of the analysis, not a constitutive parameter. Applying similar arguments to a truncated region $\{B_x : x_0 \leq x \leq R \}$ provides the linear
increase relation versus distance from the band for matrix stress $\tau_x = \tau(x_0)/(1-\phi_0)$ at any point $x_0 \in (0,\xi)$ and time $t$ in the rigid region implied in Ref.~\cite{grady1992}:  $ \tau_y - \tau_x = (1 - x_0 / \xi) (\tau_y - \tau_0)$.

%\pagebreak
\section{Shear-band governing equations}
A constitutive equation relates local shear stress, $\tau$, on the plane of the shear band 
to the slipped distance $\psi$, local temperature $\theta$, 
and rate of slip $\dot{\psi} = \d \psi / \d t$. This equation is independent from \eqref{eq:taudeg}.
A local energy balance relates temperature rise to dissipation on the slip surface
and heat transfer between the band and its environment.
Though motivated from continuum theory, these are discrete, rather than continuum,
governing equations, specialized to the discrete surface $x = 0$.

The shear-stress constitutive law is defined as follows, extending Ref.~\cite{grady1992}
to allow for degradation from local defects and stiffening from nonlinear viscosity:
\begin{align}
\label{eq:conlaw0}
\tau(\psi, \dot{\psi}, \theta) = 
(1-\phi_0) \tau_0 (\psi, \dot{\psi}, \theta)
= (1-\phi_0)  (1 - \alpha \Delta \theta - \Lambda \psi / \psi_c) \tau_y + (1-\phi_0) (\dot{\psi}/a)^2 \eta_e.
\end{align}
Linear thermal softening parameter $\alpha  > 0 $ is defined in terms of melt temperature
and other physical properties later. Local temperature excursion is $\Delta \theta = \theta - \theta_y$.
Parameter $\Lambda \in [0,1)$ is a function of $\phi_0$ and local defect concentration $\Delta \phi$.
The larger the value of $\Lambda$, the more rapid the loss of strength in the band due to geometric or
damage-softening from these concentrated defects. As $\Lambda \rightarrow 1$, the material becomes more brittle.
Specifically,
a degradation factor $\lambda$ is of the form
\begin{align}
\label{eq:conparams}
\lambda = \lambda(\Delta \phi, \phi_0) = 1 - \Lambda(\Delta \phi, \phi_0), \quad 
\lambda(0,\phi_0) = 1, \quad \lambda \in (0,1]; 
\end{align}
Noting $2 \dot{\psi}/{a}$ is the homogenized nominal shear strain rate over the width $a$ of the band, an effective quadratic viscosity coefficient is $\eta_e/4$. This material parameter can depend on microstructure (e.g., defect concentration), temperature, and as defined later, the relative strain rate $\dot \gamma$ of the background continuum. 
Motivation for nonlinear viscosity of the shearing interface was discussed in Section 1.
Dividing \eqref{eq:conlaw0} by $1 -\phi_0$,
\begin{align}
\label{eq:conlaw}
\tau_0 = \tau_y (1 - \alpha \Delta \theta - (1- \lambda) \psi / \psi_c) + \eta_e (\dot{\psi}/a)^2;
\qquad \eta_e = \eta_e(\theta,\Delta \phi, \phi_0, \dot{\gamma}).
\end{align}
Eliminating $\tau_0$ using \eqref{eq:taudeg} and substituting $\psi$ and $\dot{\psi}$ from \eqref{eq:rigidsol},
temperature of the shear band is 
\begin{equation}
\label{eq:thetatime}
\theta(t) - \theta_y = \frac{1}{\alpha} \left[ \frac{\lambda}{\psi_c} {\psi} + \frac{\eta_e}{\tau_y a^2} \dot{\psi}^2 \right]
=\frac{ \tau_y \dot{\gamma}}{18 \alpha \rho_0 \psi^2_c} \left[{\lambda}  t^3 + \frac{\eta_e \dot{\gamma} }{2 \rho_0 a^2} t^4 \right] =  \frac{1}{\alpha} \left(\frac{t}{t_c}\right)^3 \left[ \lambda + \frac{\eta_e \dot{\gamma}}{2 \rho_0 a^2} t \right].
\end{equation}
A second equation for temperature is furnished by an energy balance; a discrete analog of \eqref{eq:contlaws} is
\begin{equation}
\label{eq:contlawsb}
\rho c \dot{\theta} = {\beta} \tau \frac{ 2 \dot{\psi}}{a} - \frac{2 k_e}{a^2} \Delta \theta \quad \Rightarrow \quad 
\dot{\theta} =  \frac{2 \beta}{\rho_0 c a} \tau_0 \dot{\psi} - \frac{ 2 \chi_e}{a^2} \Delta \theta;
\quad \chi_e = \chi_e (\theta, \Delta \phi, \phi_0).  
\end{equation}
Derivative $\partial^2 \theta / \partial x^2 \approx  -2 \Delta \theta / (a^2)$ \cite{grady1991,grady1992,grady1994}  can be expressed as a second-order difference formula centered at $x = 0$ having
grid spacing $a$ and vanishing $\Delta \theta$ at $x = \pm a$.
Effective diffusivity of the band surface $\chi_e = k_e / (\rho_0 c)$, enriching constant diffusivity $\chi_0$ of the homogeneous background continuum in \eqref{eq:contlawsred}, can depend on local temperature and defects.
From \eqref{eq:thetatime}, where $\theta_c$ and $\eta_c$ are temperature change and viscosity at $t_c$,
\begin{equation}
\label{eq:thetac}
\theta_c = \theta(t_c) - \theta_y = \lambda / \alpha + \eta_c \dot{\gamma} \, t_c /(2 \alpha \rho_0 a^2); \qquad \eta_c = \eta_e(\theta_c,\Delta 
\phi, \phi_0, \dot{\gamma}).
\end{equation}

As in Refs.~\cite{grady1991,grady1992,sheng2024}, the energy balance in \eqref{eq:contlawsb} is weakly enforced, in integral form, over the time spanned by band evolution.
Noting $\psi(t_c) = \psi_c$, $\d \, \Delta \theta = \dot{\theta} \d t$, $\d \psi = \dot{\psi} \d t$ and using \eqref{eq:taudeg},
\begin{align}
\label{eq:ebalint}
\theta_c = \int_0^{\theta_c} \d \theta = \frac{2 \beta}{\rho_0 c a} \int_0^{\psi_c} \tau_0 \, \d {\psi}  
- \frac{ 2 }{a^2} \int_0^{t_c} \chi_e \Delta \theta \, \d t
= \frac{ \beta \tau_y}{\rho_0 c a}  \psi_c - \frac{ 2 \lambda \chi_0 }{a^2} \int_0^{t_c} \Delta \theta \, \d t.
\end{align}
The rightmost expression assumes an implicit functional form for $\theta$ dependence of $\chi_e$: averaged over the duration of
shear banding, diffusivity degrades similarly to stress by a factor of $\lambda$.
Defects can induce cracks or tears in the material and an effective temperature rise in micro-bands between
voids, all of which serve to decrease conductivity $k_e$ and more so, viscosity $\eta_e$. The latter at $t_c$ is
\begin{align}
\label{eq:etaeff}
\eta_c (\Delta \phi, \phi_0, \dot{\gamma}) = \lambda^2 (\Delta \phi, \phi_0) {\eta \dot{\gamma}_0}/{ \dot{\gamma}} \, ;
\qquad \eta = {\rm const} \geq 0, \quad {\dot \gamma}_0 = {\rm const} > 0.
\end{align}
Normalization by  $\dot{\gamma}$ of the underlying medium  is needed to keep viscous forces reasonable at high rates; otherwise, $t_c$ and $a$ are overestimated. Integrating \eqref{eq:thetatime} and using \eqref{eq:rigidsol} and \eqref{eq:thetac}, \eqref{eq:ebalint} is
\begin{align}
\label{eq:ebalint2}
\frac{\lambda}{\alpha} \left( 1 + \frac{\lambda \eta \dot{\gamma}_0 } {2 \rho_0   a^2} t_c \right) = 
\frac{ \beta }{\rho_0 c a} \left(
 \frac{\tau^3_y \dot{\gamma}}{18 \rho_0 } t_c^3 \right)^{1/2}-  
\frac{ \lambda^2 \chi_0 }{2 \alpha a^2} t_c \left( 1 + \frac{2}{5} \frac{\lambda \eta \dot{\gamma}_0}{\rho_0 a^2} t_c \right),
\end{align}
where an assumption similar to that used for $\chi_e$ is used for integrating $\eta_e$. Rearranging,
\begin{align}
\label{eq:rearr}
\tau_y \psi_c & =  \left(
  \frac{\tau^3_y \dot{\gamma}}{18 \rho_0 } t_c^3 \right)^{1/2}   =
 \frac{\lambda \rho_0 c a }{\alpha \beta} \biggr{\{} 
 1 + \frac{t_c}{2 a^2} \left[ \lambda \chi_0 + \frac{\lambda \eta \dot{\gamma}_0}{\rho_0} \left(1
 + \frac{2}{5} \frac{\lambda \chi_0}{a^2} t_c \right) \right] \biggr{\}}
 \approx  \frac{\lambda \rho_0 c a }{\alpha \beta} \biggr{\{} 
 1 + \frac{ \Xi t_c}{2 a^2} \biggr{\}}, 
 \\
 \label{eq:Xidef}
  \Xi & = \lambda (\chi_0 + \eta \dot{\gamma}_0 / \rho_0).
 \end{align}
 Smallness of the term in $\Xi$ proportional to $\frac{2}{5}$ relative to unity,
 accounting for coupling of thermal and viscous diffusion, is verified a posteriori.
 In terms of $\psi_c$, \eqref{eq:rearr} is, with $\Gamma$ a dissipated energy,
 \begin{align}
 \label{eq:rearr2}
 \psi_c^2 = \frac{4}{9} \frac{\tau_y \dot{\gamma} a^6}{\rho_0 \Xi^3} \left( \frac{\alpha \beta \tau_y}{\lambda \rho_0 c a} \psi_c - 1\right)^3  \Leftrightarrow 
 \Gamma^2 =  \frac{\tau^3_y \dot{\gamma} a^6}{9 \rho_0 \Xi^3} \left( \frac{2 \alpha \beta }{\lambda \rho_0 c a} \Gamma - 1\right)^3; \quad \Gamma = \int_0^{\psi_c} \tau_0 \, \d \psi = \frac{1}{2} \tau_y \psi_c.
 \end{align}

%\pagebreak
\section{Analytical solution}
As posited by Grady \cite{grady1991,grady1992,grady1994}, among all possible band
widths $a_\star$, the true or optimum shear band width $a$ should be achieved with the lowest
expenditure of energy, meaning the minimum of $\Gamma$ with respect to band width.
As shown earlier by Grady and Kipp \cite{grady1987}, the same optimal band width corresponds
to a minimum of shear-band evolution time $t_c$.  If, for example, numerous bands nucleate within
a heterogeneous sample of material, stress release waves from more rapidly growing bands 
are expected to arrest more slowly growing bands. Ultimately, if the sample dimensions are not
too large, a single shear band dominates the failure process. This is the case witnessed in
many SHPB torsion experiments \cite{cho1990,duffy1992,fellows2001} and simulations \cite{vishnu2022}, wherein specimen failure is linked to a single primary band. 
Denoting $\Gamma_\star = \Gamma(a_\star)$, differentiating the second of \eqref{eq:rearr2} gives
\begin{align}
\label{eq:dGammada}
3 \Gamma_\star \frac{ \d \Gamma_\star }{ \d a_\star} & = 
 \frac{\tau^3_y \dot{\gamma} a_\star^5}{ \rho_0 \Xi^3} \left( \frac{2 \alpha \beta }{\lambda \rho_0 c a_\star} \Gamma_\star - 1\right)^3
 +  \frac{\alpha \beta \tau^3_y \dot{\gamma} a_\star^5}{ \lambda c \rho^2_0 \Xi^3}
 \left( \frac{2 \alpha \beta }{\lambda \rho_0 c a_\star} \Gamma_\star - 1\right)^2
   \left(  \frac{ \d \Gamma_\star}{\d a_\star}
 - \frac{\Gamma_\star}{a_\star}   \right).
\end{align}
Redefining $(a, \Gamma)$ as $(a_\star, \Gamma_\star)$ for which $\d \Gamma_\star / \d a_\star = 0$,
\eqref{eq:dGammada} gives the non-degenerate ($\Gamma \neq 0$) relation
%\begin{equation}
%\label{eq:ndegen}
$\alpha \beta \Gamma = \lambda \rho_0 c a$.
%\end{equation} 
Substituting into \eqref{eq:rearr2}, optimum shear band width and %matrix failure 
energy per unit area are
\begin{equation}
\label{eq:opt}
a = \left( \frac{9 \lambda^2 c^2 \rho_0^3 \Xi^3}{\alpha^2 \beta^2 \tau_y^3 \dot{\gamma}} \right)^{1/4}, \qquad \Gamma = 
\frac{\lambda \rho_0 c}{\alpha \beta} \left( \frac{9 \lambda^2 c^2 \rho_0^3 \Xi^3}{\alpha^2 \beta^2 \tau_y^3 \dot{\gamma}} \right)^{1/4}.
\end{equation}
The shear band width in \eqref{eq:opt} can be compared with Grady's original inviscid solution \cite{grady1991,grady1992}:
\begin{equation}
\label{eq:aG}
\{ \lambda \rightarrow 1, \beta \rightarrow 1, \Xi \rightarrow \chi_0 \} 
\quad \Rightarrow \quad a  \rightarrow [9 c^2 \rho_0^3 \chi_0^3/({\alpha^2 \tau_y^3 \dot{\gamma}})]^{1/4}.
\end{equation}
As in \eqref{eq:aG}, in the limit $\Delta \phi \rightarrow 0 \Rightarrow \lambda \rightarrow 1$, localization initiates
from a local perturbation in properties or conditions too small to measurably affect \eqref{eq:conlaw0} or \eqref{eq:contlawsb}.

Neither the linearly viscous augmentation explored briefly in Ref.~\cite{grady1992} nor
the viscoplastic modification in Ref.~\cite{sheng2024} retain the simplicity of the original solution
as maintained here in \eqref{eq:opt}. Linear viscosity \cite{grady1992} does not seem to allow a closed-form analytical solution,
and prior extensions \cite{grady1992,sheng2024}, like \eqref{eq:aG}, produce a singular solution
$a \rightarrow 0$ and $\Gamma \rightarrow 0$ as $\chi_0 \rightarrow 0$. In contrast, \eqref{eq:opt} with $\Xi$ defined
in \eqref{eq:Xidef}, allows for nonzero $a$ and $\Gamma$ even in the limit of null conduction.
This implies other microscopic dissipative mechanisms, distinct from heat flow, can preclude a near zero-width, zero-energy band.
Ref.~\cite{sheng2024} also contains a constant $\beta$; prior analytical models did not address porosity $\phi_0$ or heterogeneity  here depicted by $\lambda$.
It can be verified graphically that \eqref{eq:opt} corresponds to a global minimum of $\Gamma_\star(a_\star)$ \cite{grady1992,sheng2024}.
As in Ref.~\cite{grady1992}, the second of \eqref{eq:rearr2} and \eqref{eq:opt} can be recast as
\begin{equation}
\label{eq:optnorm}
2 (\Gamma_\star / \Gamma)/(a_\star / a) = 1 + [(\Gamma_\star / \Gamma)^{1/3}/(a_\star / a)]^2.
\end{equation}

Substitution of \eqref{eq:opt} into governing equations of
Sections 2 and 3 produces other band characteristics. Critical shear displacement $\delta_b$, final slipped distance $\psi_c$,
localization time $t_c$, final band temperature excursion $\theta_c$,
relaxation distance function $\xi(t)$, and its speed $ \d \xi / \d t$ are
\begin{align}
\label{eq:deltab}
\delta_b & = 2 \psi_c = \frac{ 4 \Gamma}{\tau_y} = \frac{4 \lambda \rho_0 c}{\alpha \beta \tau_y} \left( \frac{9 \lambda^2 c^2 \rho_0^3 \Xi^3}{\alpha^2 \beta^2 \tau_y^3 \dot{\gamma}} \right)^{1/4} =
 \frac{4 \lambda \rho_0 c}{\alpha \beta \tau_y} a, \\
\label{eq:tcritical}
t_c & = \left( \frac{18 \rho_0 }{\tau_y \dot{\gamma}} \right)^{1/3} \psi_c^{2/3} = 6 \left( \frac{ \lambda^2 c^2 \rho_0^3 \Xi}{\alpha^2 \beta^2 \tau_y^3 \dot{\gamma}} \right)^{1/2} = \frac{2 a^2}{\Xi},
\\
\label{eq:thetacritical}
\theta_c &= \frac{\lambda}{\alpha} \left[ 1 + \frac{\eta \dot{\gamma}_0}{\rho_0 \chi_0 + \eta \dot{\gamma}_0} \right];
\qquad \xi(t) = \frac{1}{12} \frac{\tau_y^2}{\rho_0 \Gamma} t^2, \quad
\frac{\d \xi}{\d t} = \frac{1}{6} \frac{\tau_y^2}{\rho_0 \Gamma} t.
\end{align}

Assume that in the absence of localized defects (i.e., $\lambda = 1$), temperature rise in the band
is some fraction $f_M \in (0,1]$ of the ambient-pressure melt temperature $\theta_M$,
defined in what follows in terms of room temperature $\theta_0$ \cite{zhou2006} (here, $\theta_0 = 293\,$K \cite{vishnu2022}). Then 
the thermal softening coefficient
 $\alpha$ satisfies
\begin{equation}
\label{eq:alphasat}
\alpha = \frac{1}{f_M \theta_M}  \left[ 1 + \frac{\eta \dot{\gamma}_0}{\rho_0 \chi_0 + \eta \dot{\gamma}_0} \right],
\qquad 
f_M = \frac{\theta_M - \theta_0}{\theta_M}.
\end{equation}
If the effect of viscous dissipation on $\alpha$ is omitted as in Refs.~\cite{grady1992,sheng2024}, then
$\theta$ can greatly exceed the melt temperature even before localization completes, which
is physically unrealistic. Note also that $\eta$ in the present model accounts for strain-rate sensitivity of
 \textit{solid} material in the shear-band process zone, and not the viscosity of a liquid metal.
 Latent heat of melting considered in Ref.~\cite{claytonarx2025} is not included here or in Refs.~\cite{grady1992,sheng2024}. 
 From \eqref{eq:tcritical}, the higher-order term omitted in \eqref{eq:rearr} is
 \begin{equation}
 \label{eq:noncoup}
  ({2}/{5}) ({\lambda \chi_0}/{a^2}) t_c   =({4}/{5}) [{\rho_0 \chi_0}/({\rho_0 \chi_0 + \eta \dot{\gamma}_0 })].
 \end{equation}
 In most applications of the model in Section 5, this term is small relative to unity whenever $\eta > 0$.
 
 The expression for length of the ductile-failure process zone, $\zeta_b$, of Refs.~\cite{grady1992,sheng2024}, in turn based on Ref.~\cite{lawn1975}, is adapted to permit nominal porosity $\phi_0$. The elastic shear modulus of the matrix
 is $\mu = \rm{const}$, and that of the porous material is $\hat{\mu} = (1-\varphi_0) \mu$.
 Recall dynamic strength of the matrix is $\tau_y$, and that of the porous material is
 $\hat{\tau}_y = (1-\phi_0) \tau_y$. Correspondingly, dissipated energy \eqref{eq:rearr2} of the shear band scaled per unit area of \textit{porous} material is $\hat{\Gamma} = (1-\phi_0) \Gamma$.
 Extending Ref.~\cite{grady1992} and defining the average propagation speed over the kinetic history as $\upsilon_b$, porosity terms cancel and
 \begin{equation}
 \label{eq:zetab}
 \zeta_b = \frac{\pi}{4} \frac{\hat{\mu} \hat{\Gamma}}{\hat{\tau}_y^2} = \frac{\pi}{4} \frac{{\mu} {\Gamma}}{{\tau}_y^2}
 = \frac{\pi}{4} \frac{\lambda \rho_0 c \mu}{\alpha \beta \tau_y^2} \left( \frac{9 \lambda^2 c^2 \rho_0^3 \Xi^3}{\alpha^2 \beta^2 \tau_y^3 \dot{\gamma}} \right)^{1/4},
 \qquad \upsilon_b = \frac{\zeta_b}{t_c} = \frac{\pi}{8} \frac{{\mu} \Xi} {{\tau}_y^2} \frac{\Gamma}{a^2}.
 \end{equation}
 According to \eqref{eq:zetab}, a narrower band propagates faster than a wider band of matching energy $\Gamma$.
 
In Ref.~\cite{grady1987}, an optimum periodic band spacing of $2 \xi(t_c)$ was proposed for densely packed shear
bands in a shock-process zone. That expression for periodic spacing is not used here in calculations depicting SHPB experiments, wherein one dominant band leads to global stress collapse. Rather, band
spacing in the current application, at lower rates than for shock, effectively equals or exceeds the gage length $L$ of the torsion specimen. Omission of elastic waves in the rigid-plastic model (Section 2) also renders
 $2 \xi(t_c)$-depiction of spacing less accurate at lower rates \cite{grady1992}.
 
 In application of the present framework to study simple shear, the minimum strain at which localization initiates from a small perturbation is assumed to be the peak instability strain ${\gamma}_p$ \cite{wright2002,yan2021,shawki1988,anand1987}.  Assuming homogeneous adiabatic deformation up to
 this instability threshold, initial conditions for the localization analysis given in Appendix A for a power-law thermo-viscoplastic solid sheared at constant rate $\dot{\gamma}$ from ambient (e.g., room) starting temperature $\theta_0$ are
 \begin{align}
 \label{eq:ics}
 \theta_y = \theta(\gamma_p; \dot{\gamma}, \theta_0), \qquad \tau_y =  \tau(\gamma_p; \dot{\gamma}, \theta_0)/ (1-\phi_0), \qquad t_p = \gamma_p / \dot{\gamma},
 \end{align}
 where $t_p \leq t_0$ is the time at peak stress and 
  $\tau$, $\theta$, and $\gamma_p$ are given by \eqref{eq:flowstress}, \eqref{eq:thetan}, and \eqref{eq:instrain}. 
  
  Henceforth for clarity, the time datum is shifted so $t = 0$ corresponds to the start of the experiment when $\gamma = 0$, $\tau =0$, and $\theta = \theta_0$.
  The shear-band evolution model of Sections 2 and 3 spans domain $t \in [t_0,t_0 + t_c]$, where the magnitude of
  the process time is $t_c$, and $t_0 \geq t_p$.
  The failure time at which $\gamma = \gamma_c$ is $t_f = \gamma_c / \dot{\gamma} \geq t_p + t_c$.
  
  Over the period $t \in (t_p, t_0)$, the strain field in a dynamic torsion test becomes heterogeneous, prior to the
  late stage of localization.
  Shear-strain rate (but not strain itself) increases close to location of the pending shear band and necessarily decreases farther away \cite{duffy1992,cherukuri1995,fellows2001}. The
  rigid-plastic framework of Section 2 is not designed to capture this pre-localization process,
  and initiation time $t_0$ is not known a priori. Instead, define a ratio $r_0$:
  \begin{equation}
  \label{eq:ti}
   t_f = t_0 + t_c = t_p + (t_0 - t_p) + t_c = t_p + r_0 t_c, \quad r_0 = (\gamma_c - \gamma_p)/(\dot{\gamma} \, t_c) = (t_f - t_p)/ t_c \geq 1.
  \end{equation}
Net post-peak strain contribution to $\gamma_c$ for $t \in (t_p, t_c]$ from material outside the 
band is depicted as arising from a region of reduced average strain rate $\dot{\bar{\gamma}} = \dot{\gamma}/r_0$ over scaled duration $\bar{t} = t_c r_0 $ such that
\begin{align}
\label{eq:gammacrit}
\gamma_c & = \gamma_p + \varphi_b \gamma_b + (1-\varphi_b) \dot{\bar{\gamma}} \, \bar{t} = 
\gamma_p + \varphi_b \gamma_b + (1-\varphi_b) \dot{{\gamma}} \, t_c, \\
 \gamma_b & = \frac{\delta_b}{a} = \frac{4 \lambda \rho_0 c}{\alpha \beta \tau_y},  \qquad \varphi_b = \frac{a}{L}.
\end{align}
Specimen length $L$ necessarily affects $\gamma_c$. The volume fraction of banded material is $\phi_b$, and $\gamma_b$ is average
 shear strain supported by the band of fixed width $a$ (i.e., displacement redistributed linearly over the width
as in Fig.~\ref{fig1}).
This representation of $\phi_b$ assumes the band traverses the circumference of the specimen,
accelerating if necessary as $t \rightarrow t_f^+$. An equivalent result is obtained if the band is assumed to nucleate homogeneously with respect to the $y$ coordinate; then \eqref{eq:zetab} is irrelevant.
Critical strain $\gamma_c$ is conveniently independent from $r_0$, though the latter can be calculated
if $\gamma_c$ and $\gamma_p$ are known with $t_c$ found through \eqref{eq:tcritical}.
Then from \eqref{eq:ti}, $t_0 = t_p + (r_0 - 1) t_c$; as $r_0 \rightarrow 1$, $t_0 \rightarrow t_p$.

To complete the model, a specific form of $\lambda(\Delta \phi, \phi_0)$ of \eqref{eq:conparams}
is prescribed.  For the application in Section 5, an exponential form proves useful
to describe numerical data \cite{vishnu2022}:
\begin{equation}
\label{eq:lambdadef}
\lambda = \exp \left[ - \omega \frac{\Delta \phi}{\phi_0} \right], \qquad \omega = {\rm const} \geq 0.
\end{equation}
Argument $\Delta \phi / \phi_0$ is calculated from porosity distribution data via procedures of Appendix B.

%\pagebreak
\section{Results}

\subsection{Parameterization}

The framework of Sections 2 through 4 is first exercised to describe dynamic finite element (FE) simulations of
porous polycrystalline metals reported by Vishnu et al.~\cite{vishnu2022}. In these simulations, thin-walled
tubes were subjected to dynamic torsion. Loading conditions mimicked SHPB experiments, albeit with
periodic rather than finite boundaries to minimize edge effects and allow shear banding to occur anywhere
within the gage length.
Two metals were studied: commercially pure titanium (Ti) and HY-100 steel. Constitutive models of the general form
in Appendix A were used: power-law strain hardening, strain-rate hardening, and thermal softening.
Simulations applied strain rates $\dot{\gamma}$ from $10^2$/s to $10^4$/s, with a majority of results
presented for $\dot{\gamma} = 10^3$/s. The internal diameter, thickness, and length $L$ of the tube were
9.5, 0.38, and 2.5 mm, respectively. Solid material was idealized as isotropic, but discrete voids were resolved in
FE meshes. Representative microstructures were reconstructed based on experimental characterization of AM metals
\cite{marvi2021}. Average porosities for microstructures modeled predominantly in Ref.~\cite{vishnu2022}
and depicted in the current study are small, with $\phi_0$ on the order of $10^{-3}$ or 0.1\%.
Simulation of a fully dense microstructure ($\phi_0 =0$) was also reported for Ti.
A benefit of comparison versus FE results is detailed information on many microstructures
is available \cite{vishnu2022}. Drawbacks are results are subject to
limitations of underlying constitutive models, parameters, and numerical (e.g., mesh) resolution.

Heat conduction was omitted in Ref.~\cite{vishnu2022}, but a prior study with dynamic compression loading conditions \cite{vishnu2022b}
showed very small effects of heat conduction on localization behavior. Shear banding was promoted primarily
by pores and thermal softening and was modulated by strain rate sensitivity. Localized flow tended to initiate at the
largest pores or regions of largest local void volume fraction, with secondary localizations at smaller defects. 
As deformation progressed, the dominant shear
band tended to emerge over the circumference and override effects of secondary bands, leading to final
failure. This concurs with the justification in Refs.~\cite{grady1987,grady1992}, noted in Section 4, for the dominant 
shear band to follow a minimum critical-time or minimum surface-energy principle.
Critical strain $\gamma_c$ at pending specimen failure tended to increase weakly with $\phi_0$ and more
strongly with largest pore size. In the present framework, effects of large voids or clusters of voids are
modeled in a continuum-distributional sense through local concentration function $\Delta \phi$.
The ratio $\Delta \phi / \phi_0$ measures the effective increase in void fraction in the vicinity of the largest defect relative
to the global average $\phi_0$. A mathematical definition and procedure by which defect concentration
$\Delta \phi / \phi_0$ is calculated from the microstructure data of Ref.~\cite{vishnu2022} are given in Appendix B.
This concentration, in turn, is used in \eqref{eq:lambdadef} to compute softening function $\lambda = 1 - \Lambda$, recalling
$\lambda = 1$ corresponds to a homogeneous material. Numerical results \cite{vishnu2022} showed a modest
decrease in $\gamma_c$ when strain rate was increased from $10^3$ to $10^4$/s. For the fully
dense material, a single dominant shear band formed in the center of the specimen due to symmetry. The band in
this instance appeared to be wider (i.e., larger $a$) and delayed (larger $t_c$ and $\gamma_c$) relative to 
porous microstructures.

Parameters entering the current analysis are given in Table~\ref{table1} with supporting references.
Far-field conditions, namely $\tau_y$, $\theta_y$, and $\gamma_p$, are found from \eqref{eq:ics} and
procedures in Appendix A. Regarding the latter, parameters used in \eqref{eq:flowstress} are converted
to shear-stress versus shear-strain space
from Ref.~\cite{vishnu2022}, latter couched in terms of effective (i.e., Von Mises) stress and strain.
The viscoplasticity parameters influence $\tau_y$, $\theta_y$, and $\gamma_p$, but do not otherwise affect
the analysis.  Only two parameters require calibration: the viscosity product $\eta \, \dot{\gamma}_0$
and the softening parameter $\omega$ entering \eqref{eq:lambdadef}.
The former ($\eta$) is tuned for each material to match values of $\gamma_c$ in the limit of zero or very small $\Delta \phi$
from simulations at $\dot{\gamma}= \dot{\gamma}_0 = 10^3$/s \cite{vishnu2022}. 
The latter ($\omega$) is fit by regression of $\gamma_c$ to simulation data \cite{vishnu2022} over a range of maximum pore sizes and
average porosities (Appendix B). Outcomes are discussed in Section 5.2.
Recall $\gamma_c$ is predicted by the current approach via \eqref{eq:gammacrit}.

In most results reported subsequently, experimental values of thermal diffusivity $\chi_0$ are used
for realism, even though simulations \cite{vishnu2022} assume locally adiabatic conditions.
In the present approach, thermal diffusivity can be eliminated by increasing viscosity
$\eta$. If $\eta_0$ is a value of $\eta$ with conduction enabled, then a default
``artificial'' viscosity to offset $\chi_0$ can be included in $\Xi$ of \eqref{eq:Xidef}:
\begin{equation}
\label{eq:default}
\eta \rightarrow \eta_0 + \rho_0 \chi_0/ \dot{\gamma}_0, \qquad
\Xi \rightarrow \lambda \eta \dot{\gamma}_0 / \rho_0.
\end{equation}
As shown in Section 5.2, results of equal accuracy can be obtained via \eqref{eq:default} or $\chi_0 > 0$
in \eqref{eq:Xidef}.

\begin{table}%[ht!]
\footnotesize
\caption{Properties or parameters used in calculations for titanium and HY-100 steel}
\label{table1}       % Give a unique label
\centering
\begin{tabular}{lcccl}
\hline\noalign{\smallskip}
Quantity [units]  & Rate $\dot{\gamma}$ & Value, titanium & Value, steel & Definition [reference]  \\
\noalign{\smallskip}\hline\noalign{\smallskip}
$\rho_0$ [g/cm$^3$] & $-$ & 4.51 & 7.86 & mass density (non-porous) \cite{vishnu2022} \\
$c$ [J/kg K] & $-$ & 528 & 473 & specific heat per unit mass \cite{vishnu2022} \\
$k_0$ [W/m K] & $-$ & 19 & 39 & thermal conductivity \cite{vishnu2022b,norkett2023} \\ 
$\theta_M$ [K] & $-$ & 1941 & 1793 & melt temperature \cite{norkett2023}\\ % ishikawa2012 
$\beta$ & $-$ & 0.9 & 0.9 & Taylor-Quinney factor \cite{vishnu2022,vishnu2022b} \\ 
$\mu$ [GPa] & $-$ & 43.3 & 76.0 & elastic shear modulus \cite{vishnu2022} \\
$\tau_y$ [MPa] & $10^3$/s & 631 & 629 & peak flow stress (calculated \cite{vishnu2022,claytonarx2025}) \\
 & $10^4$/s & 674 & 645 & \\
$\theta_y$ [K] & $10^3$/s & 327 & 338 & peak temperature (calculated \cite{vishnu2022,claytonarx2025}) \\
 & $10^4$/s & 327 & 338 & \\
 $\gamma_p$ [-] & $10^3$/s & 0.16 & 0.32 & peak applied strain (calculated \cite{vishnu2022,claytonarx2025}) \\
 & $10^4$/s & 0.15 & 0.31 & \\
  $\eta \cdot \dot{\gamma}_0 $ [Pa s] & $-$ & 1.22  & 0.65 & shear-band viscosity (calibrated, $\dot{\gamma}_0 = 10^3$/s) \\
 $\omega$ [-] & $-$ & 0.025 & 0.025 & defect-softening parameter (calibrated) \\
\noalign{\smallskip}\hline
\end{tabular}
\end{table}

\subsection{Comparison with data}

\begin{figure}%[ht!]
\begin{center}
 \subfigure[titanium, $10^3$/s]{\includegraphics[width=0.32\textwidth]{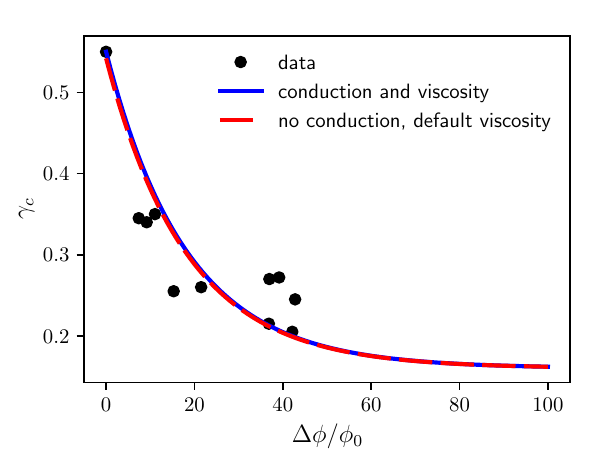} \label{fig2a}} \qquad
 \subfigure[steel, $10^3$/s]{\includegraphics[width=0.32\textwidth]{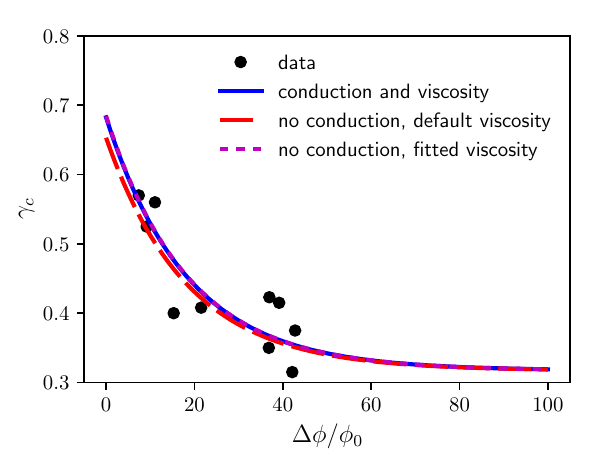}\label{fig2b}} \\
 %  \vspace{-0.2cm}
  \subfigure[titanium, two strain rates]{\includegraphics[width=0.32\textwidth]{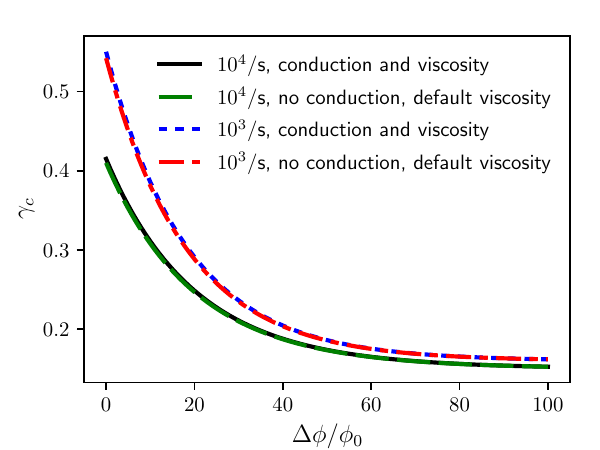}\label{fig2c}} \qquad
    \subfigure[steel, two strain rates]{\includegraphics[width=0.32\textwidth]{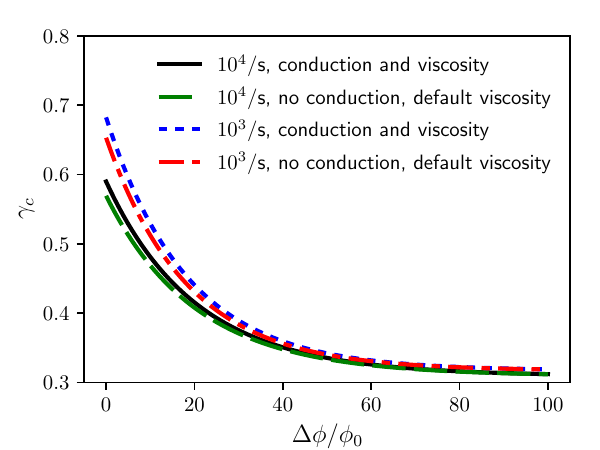}\label{fig2d}} 
 \end{center}
  \vspace{-0.5cm}
\caption{Critical average shear strain $\gamma_c$ versus normalized defect concentration
$\Delta \phi / \phi_0$:
 (a) model with and without heat conduction for titanium at $\dot{\gamma} = 10^3$/s vs.~numerical simulation data \cite{vishnu2022}
 (b) model for HY-100 steel at $\dot{\gamma} = 10^3$/s vs.~data \cite{vishnu2022} 
 (c) model for titanium at $\dot{\gamma} = 10^3$/s and $\dot{\gamma} = 10^4$/s
 (d) model for HY-100 steel at $\dot{\gamma} = 10^3$/s and $\dot{\gamma} = 10^4$/s.
 When heat conduction is omitted, default viscosity is found from \eqref{eq:default}. Fitted viscosity in (b) is tuned to 1.13$\times$ default from \eqref{eq:default} to match result
 with conduction enabled.}
\label{fig2}       
\end{figure}

Results of the analytical framework are first compared with simulation data of Vishnu et al.~\cite{vishnu2022}.
In Fig.~\ref{fig2a}, critical strain $\gamma_c$ versus normalized defect concentration $\Delta \phi / \phi_0$
provides equally satisfactory regression to the data for titanium when $\chi_0$ is enabled or when \eqref{eq:default} is used.
For steel in Fig.~\ref{fig2b}, results with \eqref{eq:default} are close but not identical to those with nonzero $\chi_0$.
However, the default value of \eqref{eq:default} can be increased by 13\% to give near-perfect agreement.
Simulation data in Ref.~\cite{vishnu2022} do not include $\gamma_c$ for multiple microstructures at loading
rates $\dot{\gamma}$ other than $10^3$/s. However, outcomes of the current analysis in Figs.~\ref{fig2c} and
\ref{fig2d} show
a slight decrease in $\gamma_c$ as strain rate increases from $10^3$ to $10^4$/s, which is consistent
with general trends reported in Ref.~\cite{vishnu2022}.
Data are obtained from simulations \cite{vishnu2022} on the same 10 porous microstructures (i.e., same
distributions of $\Delta \phi$ and $\phi_0$) but with different constitutive
parameters for Ti and HY-100 steel. Thus, if $\omega$ is only associated with microstructure heterogeneity, then
one might expect the same value to apply regardless of properties of the fully dense homogeneous solid.
This proposition is confirmed in Table 1: the same value $\omega = 0.025$ provides satisfactory descriptions of both 
materials. Note also that the value of $\eta$ for Ti is around $1.9 \times$ that of HY-100 steel.
This is notionally consistent with a higher macroscopic rate sensitivity $m$ for Ti around $2.8 \times$ that of this steel \cite{vishnu2022}.

\begin{figure}%[ht!]
\begin{center}
 \subfigure[critical strain past peak]{\includegraphics[width=0.32\textwidth]{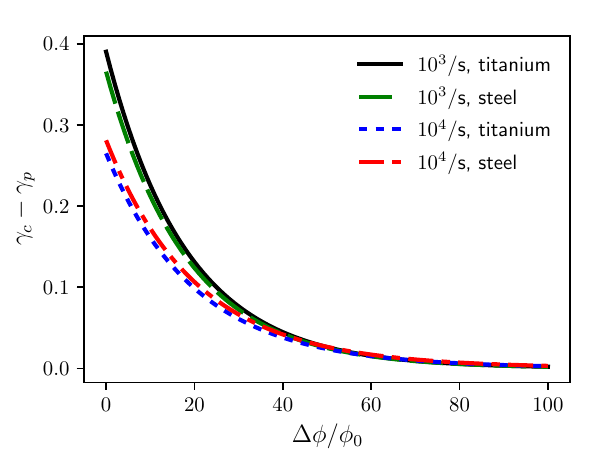} \label{fig3a}} \qquad
 \subfigure[band strain]{\includegraphics[width=0.32\textwidth]{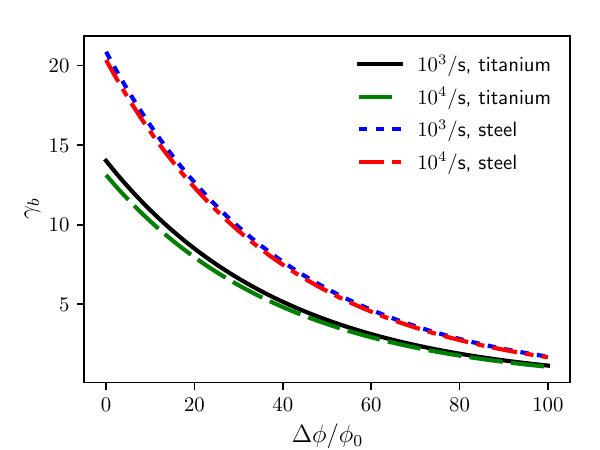}\label{fig3b}} \\
  \vspace{-0.2cm}
  \subfigure[band width]{\includegraphics[width=0.32\textwidth]{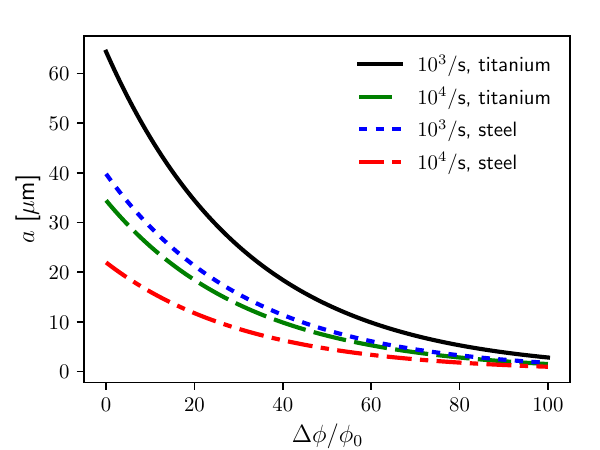}\label{fig3c}} \qquad
    \subfigure[band slip]{\includegraphics[width=0.32\textwidth]{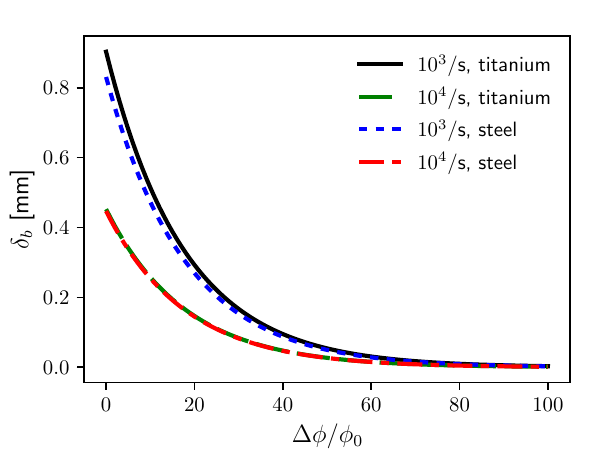}\label{fig3d}} \\
      \vspace{-0.2cm}
  \subfigure[zone length]{\includegraphics[width=0.32\textwidth]{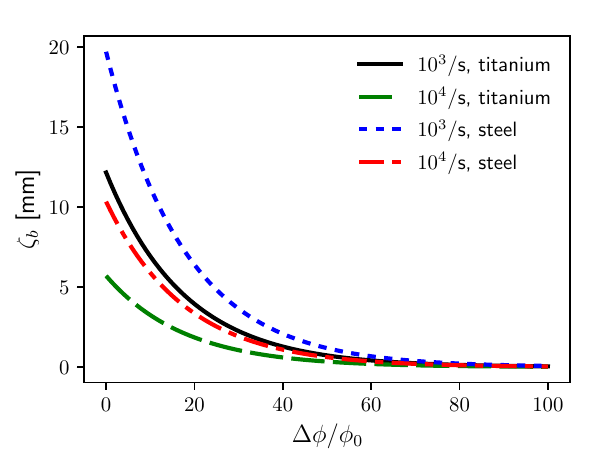}\label{fig3e}} \qquad
    \subfigure[growth time]{\includegraphics[width=0.32\textwidth]{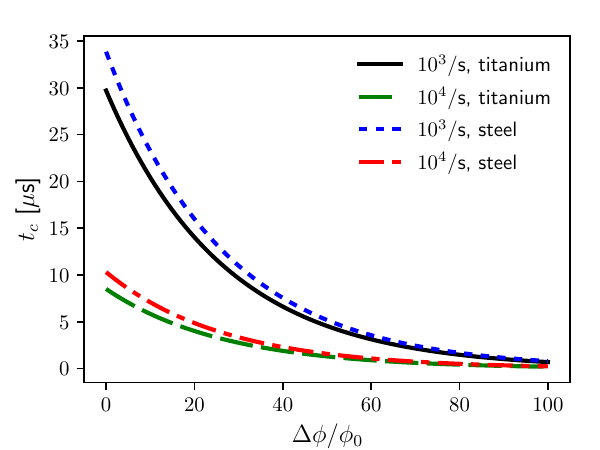}\label{fig3f}} \\
     \vspace{-0.2cm}  
      \subfigure[homologous temperature]{\includegraphics[width=0.32\textwidth]{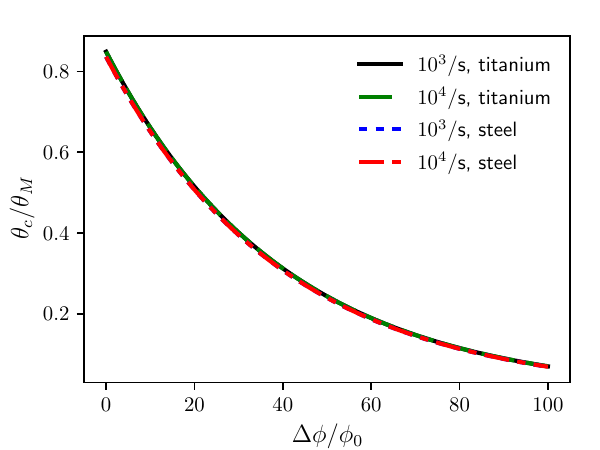}\label{fig3g}} \qquad
    \subfigure[band energy]{\includegraphics[width=0.32\textwidth]{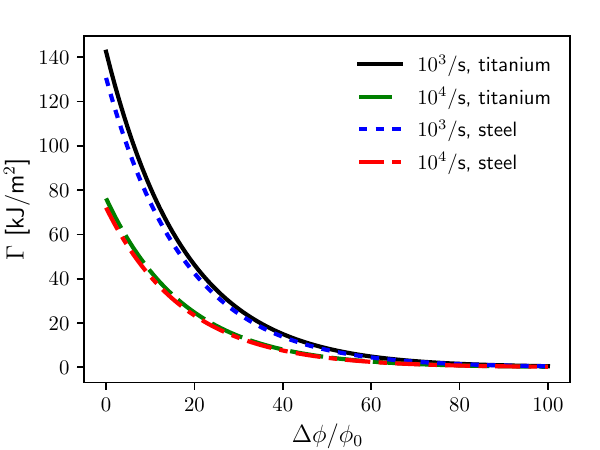}\label{fig3h}}
  \end{center}
  \vspace{-0.5cm}
\caption{Predicted influence of defect (excess pore) concentration $\Delta \phi / \phi_0$ on shear-band
characteristics of titanium and HY-100 steel using properties of Table~1:%~\ref{table1}:
(a) $\gamma_c - \gamma_p$ 
(b) $\gamma_b$
(c) $a$
(d) $\delta_b$
(e) $\zeta_b$
(f) $t_c$
(g) $\theta_c$
(h) $\Gamma$
}
\label{fig3}       
\end{figure}

Other quantities predicted by the present analytical solution of Section 4, using properties of Ti and HY-100 steel of Table~\ref{table1},
are shown in Fig.~\ref{fig3} for the range of defect concentration $\Delta \phi / \phi_0 \in [0,100]$
and strain rates $\dot{\gamma}$ of $10^3$ and $10^4$/s.  All quantities decrease with increasing local defect fraction, some more severely than others. Consistent with a 1-D spatial analysis \cite{grady1991,grady1992}, 
for a given material, loading rate, and defect concentration, 
$\zeta_b$ is an order of magnitude larger than $\delta_b$, the latter in turn an order of magnitude larger than $a$. Ratio $\theta_c/\theta_M$ decreases from unity with increasing $\Delta \phi / \phi_0$ nearly identically for both materials and loading
rates. This is a consequence of the definition of $\alpha$ in \eqref{eq:alphasat}, comparable $\theta_M$ for each material, 
and $f_M$ similar to Ref.~\cite{grady1994}.

Next, predictions of the present model are compared with experimental SHPB findings, specifically test 208 of Ref.~\cite{duffy1992}
on HY-100 steel at $\dot{\gamma} = 1350$/s.
Properties are unchanged from Table~\ref{table1} apart from 
$\tau_y = 631$ MPa and a slight reduction in $\gamma_p$ recalculated at this strain rate
via methods of Appendix A. The value of $\lambda$ is chosen such that $\gamma_c$ perfectly
matches the experimental failure strain. This steel is not described as porous in Ref.~\cite{duffy1992}, but
like any real material, heterogeneities in the microstructure that could initiate localization 
are inevitable. These may include groups of grains preferentially oriented for geometric softening, weakened grain boundaries or inclusions, or regions of lower dislocation density. Torsion specimens are also prone to thickness
variations whose effects could manifest in a lower shearing resistance \cite{molinari1987,claytonJMPS2024} via $\lambda < 1$, though the boundary value problem set-up in Section 2 does not explicitly address these irregularities.

Results in Table~\ref{table2} compare peak stress $\tau_y$, critical strain $\gamma_c$ of \eqref{eq:gammacrit}, shear band strain
$\gamma_b$, shear band width $a$, and maximum band temperature rise $\theta_c$ among
experiment \cite{duffy1992}, the present model, and reduction of the latter to Grady's original
theory \cite{grady1992,grady1994} (which did not include \eqref{eq:gammacrit}).
The present work closely matches experimental value of $\tau_y$ and predicts the final shear band
width within 40\%. Shear band strain $\gamma_b$ is over-predicted by 67\%, temperature
minimally by only 31 K. The experimental temperature measurement is prone to great uncertainty as discussed in
Ref.~\cite{duffy1992}, but the upper bound was stated as significantly lower than the melt temperature (1793 K).

The original inviscid theory of Refs.~\cite{grady1992,grady1994} gives predictions of all quantities that are less accurate,
often significantly so. Shear band strain is 3.6 times the experimental result, and width under half the experimental result. Temperature rise is excessive. An even higher, and thus even more unrealistic, temperature rise would
be predicted from the linearly viscous extension in Ref.~\cite{grady1992} and the viscoplastic analysis of
Ref.~\cite{sheng2024}. In those works, viscous dissipation increases temperature beyond that of
the original theory \cite{grady1991} (here, 1500 K with $f_M$ of \eqref{eq:alphasat}) because $\alpha \approx 1 / (\theta_M - \theta_0)$ is used therein \cite{grady1992,sheng2024}.
In the current theory, it is possible to achieve a lower $\theta_c$ than reported
in Table~\ref{table2} by decreasing $f_M$. However, given uncertainty in experimental thermal data,
this exercise, which requires recalibration of other parameters ($\eta, \omega$) for consistency, is not pursued further.

\begin{table}%[ht!]
\footnotesize
\caption{Results comparison for HY-100 steel, $\dot{{\gamma}} = 1350$/s}
\label{table2}       % Give a unique label
\centering
\begin{tabular}{lccccc}
\hline\noalign{\smallskip}
Model or experiment  & $\tau_y$ [MPa] & $\gamma_c$ & $\gamma_b$ & $a$ [$\mu$m] & $\theta_c$ [K]  \\
\noalign{\smallskip}\hline\noalign{\smallskip}
Experiment \cite{duffy1992} & 635 & 0.530 & 10 & 20$^\dagger$ &  575--1170  
 \\
Present ($\beta = 0.9$, $\lambda = 0.8$, $\eta = 6.5 \times 10^{-4}$ kg/m) & 631 & 0.530 & 16.7 & 27.9  & 1201 \\
Grady model \cite{grady1992,grady1994} ($\beta = 1$, $\lambda = 1$, $\eta = 0$) & 625 & 0.437 & 35.7 & 9.4  & 1500 \\
\noalign{\smallskip}\hline
$^\dagger$ reported width at end of experiment
\end{tabular}
\end{table}

Experimental and calculated shear band widths are compared with several other analytical predictions in 
Table~\ref{table3}, for the material and loading conditions matching Table~\ref{table2}.
Shear band widths $a$ predicted by models of Wright and Ockendon \cite{wright1992} and Dinzart and Molinari \cite{dinzart1998}, respectively, for small strain-rate sensitivity exponent $m$, are as follows:
\begin{equation}
\label{eq:alit}
a = \frac{2m}{1-m} \frac{k_0 \theta_0}{\alpha_0 g_0 \upsilon_0} \quad [{\rm{Wright \, \& \,Ockendon}}];
\qquad
a = 6 \sqrt{2}  m \frac{k_0 \theta_0}{\alpha_0 g_0 \upsilon_0} \quad [{\rm{Dinzart \, \& \,Molinari}}].
\end{equation}
Here, $\upsilon_0 = \dot{\gamma} \, h = \dot{\gamma} L / 2$ is applied velocity, $h$ is 
specimen half-width normal to the shear plane, $g_0$ is initial yield strength, and $\alpha_0$ is a dimensionless  thermal softening parameter. The flow rule \cite{dinzart1998} is $\tau = g_0(1 -\alpha_0 \theta / \theta_0)(\dot{\gamma}/\dot{\gamma}_0)^m$. Values used in the calculation of \eqref{eq:alit} for HY-100 steel from Ref.~\cite{dinzart1998}
reproduced in Table~\ref{table3} are $\upsilon_0 = 1.68$ m/s, $g_0 = 500$ MPa, $m = 0.012$, $k_0 = 54$ W/m K, $\alpha_0 = 0.248$, and $\theta_0 = 300$ K.
As witnessed in Table~\ref{table3}, the present model provides the closest result to the experimental value of $a$.
Another model due to Dodd and Bai \cite{dodd1989} supplies a formula for $a$; however, this
formula requires a priori knowledge of strain rate and temperature inside the band, rather than initial or far-field
values, so cannot be used without further assumptions on the state the band.
Formulae of Ref.~\cite{sheng2024} are not pursued due to complexity and several uncertain parameters.

\begin{table}%[ht!]
\footnotesize
\caption{Shear band width comparison for HY-100 steel, $\dot{{\gamma}} = 1350$/s, $L = 2h = 2.5$ mm }
\label{table3}       % Give a unique label
\centering
\begin{tabular}{lccr}
\hline\noalign{\smallskip}
Model or experiment  &  eqn.~\# & $a$ [$\mu$m]   & error [\%] \\
\noalign{\smallskip}\hline\noalign{\smallskip}
Experiment \cite{duffy1992} & $-$ & 20 & $-$ \\
Present model & \eqref{eq:opt} &  27.9 & $+$39.5 \\
Grady model \cite{grady1992} & \eqref{eq:aG} & 9.4 & $-$53.0 \\
Dinzart-Molinari model \cite{dinzart1998} & \eqref{eq:alit}  & 8.1 & $-$59.5\\
Wright-Ockendon model \cite{wright1992} & \eqref{eq:alit} & 1.9 & $-$90.5\\
\noalign{\smallskip}\hline
\end{tabular}
\end{table}

\subsection{Effects of viscosity and defects}
The two most novel aspects of the present theory are nonlinear viscosity and consideration of localized defects.
These manifest, respectively, through parameters $\eta$ and $\omega$.
Keeping all other parameters fixed for those of HY-100 steel in Table~\ref{table1}, effects of
ranges of $\eta$ and $\omega$ are explored next. Strain rates $\dot{\gamma}$ of $10^3$ and $10^4$/s and
defect concentrations $\Delta \phi/ \phi_0$ of 10 and 40 are considered.

\begin{figure}%[ht!]
\begin{center}
 \subfigure[critical strain]{\includegraphics[width=0.32\textwidth]{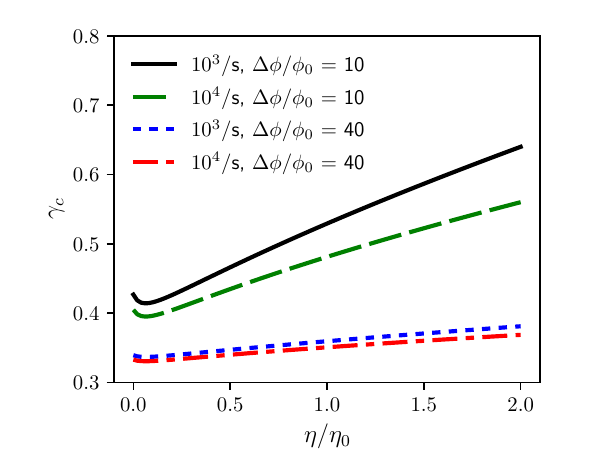} \label{fig4a}} \qquad
 \subfigure[band strain]{\includegraphics[width=0.32\textwidth]{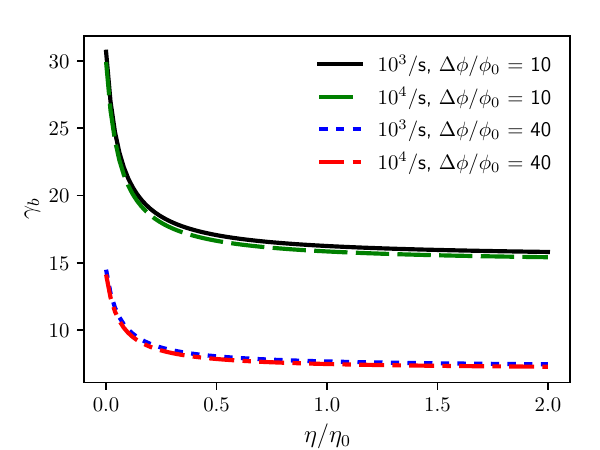}\label{fig4b}} \\
  \vspace{-0.2cm}
  \subfigure[band width]{\includegraphics[width=0.32\textwidth]{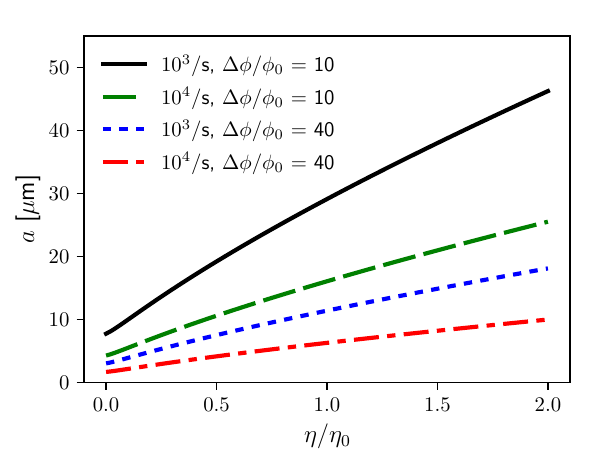}\label{fig4c}} \qquad
    \subfigure[band slip]{\includegraphics[width=0.32\textwidth]{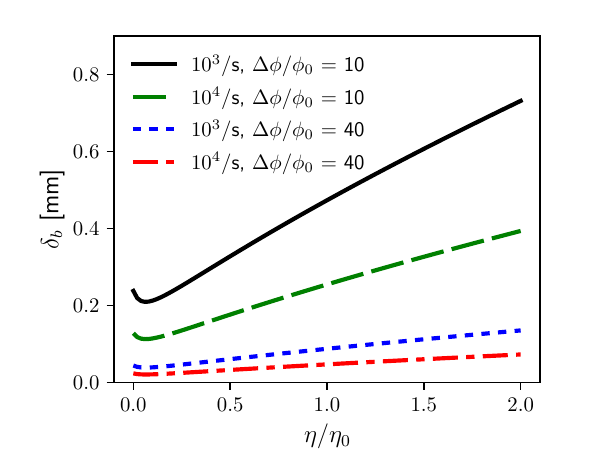}\label{fig4d}} \\
      \vspace{-0.2cm}
  \subfigure[zone length]{\includegraphics[width=0.32\textwidth]{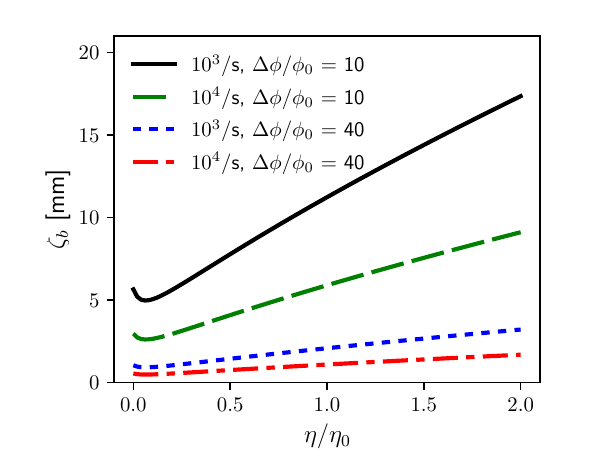}\label{fig4e}} \qquad
    \subfigure[growth time]{\includegraphics[width=0.32\textwidth]{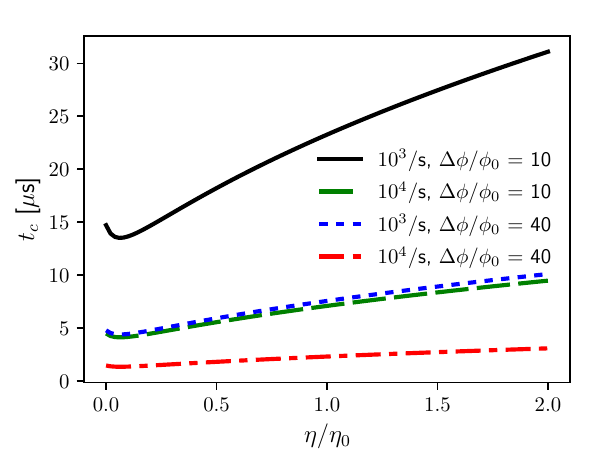}\label{fig4f}} \\
     \vspace{-0.2cm}  
      \subfigure[band temperature]{\includegraphics[width=0.32\textwidth]{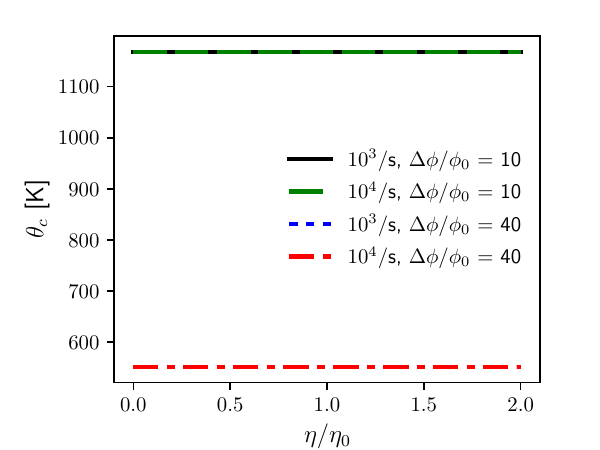}\label{fig4g}} \qquad
    \subfigure[band energy]{\includegraphics[width=0.32\textwidth]{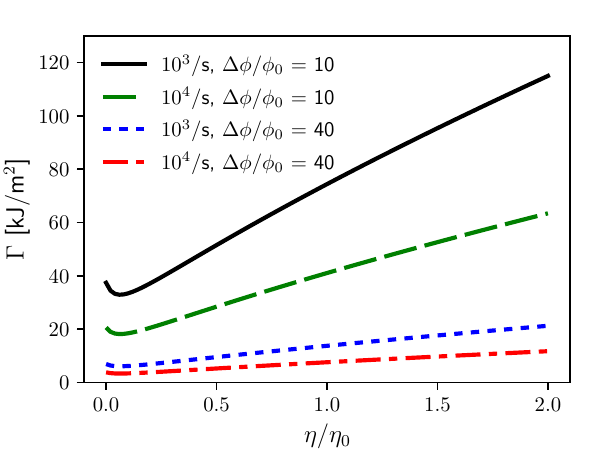}\label{fig4h}}
  \end{center}
  \vspace{-0.5cm}
\caption{Influence of viscosity $\eta$ on shear-band characteristics, two defect concentrations $\Delta \phi / \phi_0$, $\eta_0 = 4 \times 10^{-4}$kg/m, other properties for HY-100 steel of Table~\ref{table1}:
(a) $\gamma_c$ 
(b) $\gamma_b$
(c) $a$
(d) $\delta_b$
(e) $\zeta_b$
(f) $t_c$
(g) $\theta_c$
(h) $\Gamma$
}
\label{fig4}       
\end{figure}

Outcomes for $\eta / \eta_0 \in [0,2]$ at fixed $\omega = \omega_0$, where $(\eta_0,\omega_0)$ are nominal values in Table~\ref{table1}, are shown
in Fig.~\ref{fig4}. In Fig.~\ref{fig4a}, a slight decrease in $\gamma_c$ occurs with increasing $\eta$ at small $\eta/\eta_0$, followed
by a gradual increase thereafter. Effects of $\eta$ are more pronounced at smaller $\Delta \phi / \phi_0$.
With increasing viscosity, shear band strain $\gamma_b$ decreases sharply initially, then slowly plateaus in Fig.~\ref{fig4b}.
The shear band width increases in a nearly linear fashion with $\eta/ \eta_0$ in Fig.~\ref{fig4c};
the slope decreases with increasing applied strain rate and increasing defect content.
The magnitude of slip in the band, length of the process zone, and surface energy of the band
in respective Figs.~\ref{fig4d}, \ref{fig4e}, and \ref{fig4h} follow similar trends to the critical strain $\gamma_c$.
Critical time $t_c$ shows similar non-monotonicity, but the effect of strain rate is stronger.
Temperature rise in Fig.~\ref{fig4g} is independent of $\eta$ and $\dot{\gamma}$, but it is lower
in the more defective, and thus more brittle, material ($\Delta \phi / \phi_0 = 40$). A similar result was found in Ref.~\cite{claytonarx2025}
using a completely different analysis of torsion of a different steel, wherein softening from damage and fracture precluded extreme temperature rise in the band.

\begin{figure}%[ht!]
\begin{center}
 \subfigure[critical strain]{\includegraphics[width=0.32\textwidth]{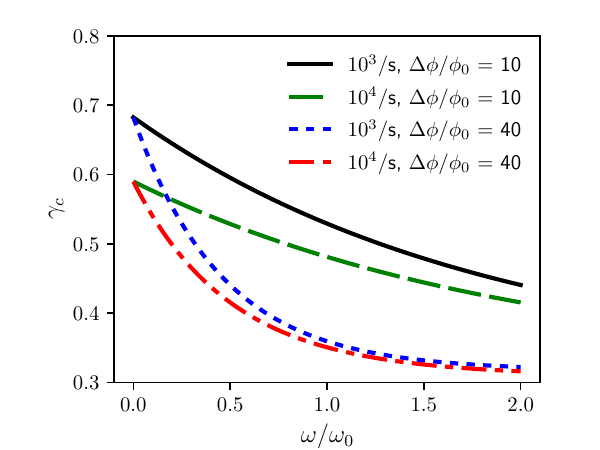} \label{fig5a}} \qquad
 \subfigure[band strain]{\includegraphics[width=0.32\textwidth]{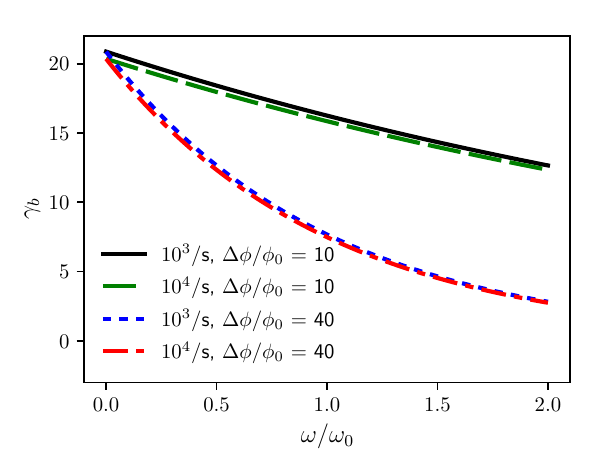}\label{fig5b}} \\
  \vspace{-0.2cm}
  \subfigure[band width]{\includegraphics[width=0.32\textwidth]{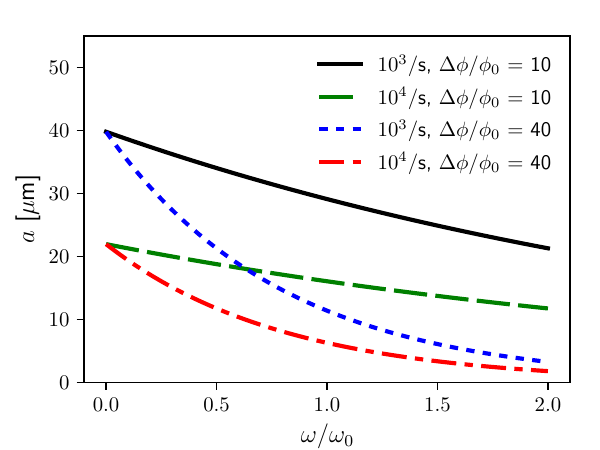}\label{fig5c}} \qquad
    \subfigure[band slip]{\includegraphics[width=0.32\textwidth]{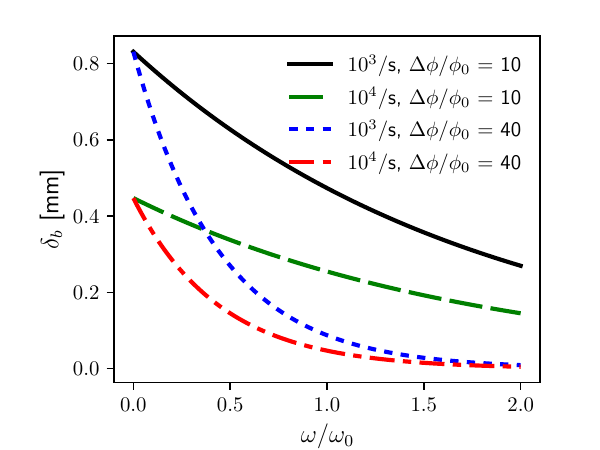}\label{fig5d}} \\
      \vspace{-0.2cm}
  \subfigure[zone length]{\includegraphics[width=0.32\textwidth]{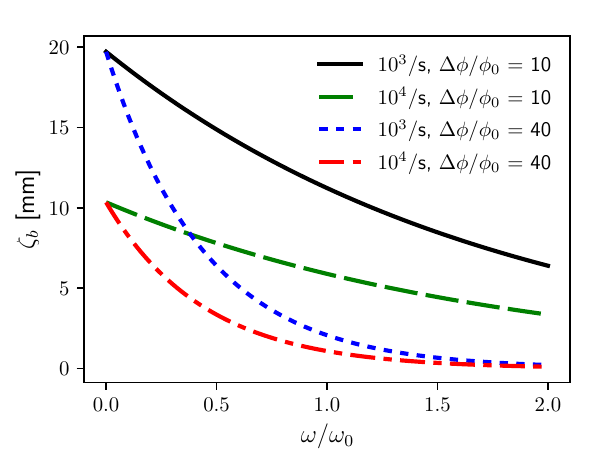}\label{fig5e}} \qquad
    \subfigure[growth time]{\includegraphics[width=0.32\textwidth]{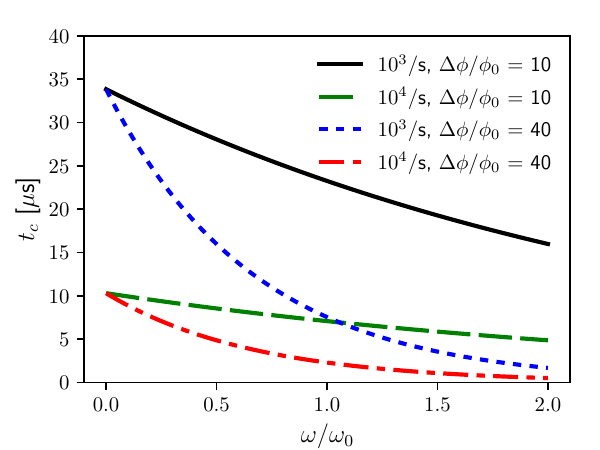}\label{fig5f}} \\
     \vspace{-0.2cm}  
      \subfigure[band temperature]{\includegraphics[width=0.32\textwidth]{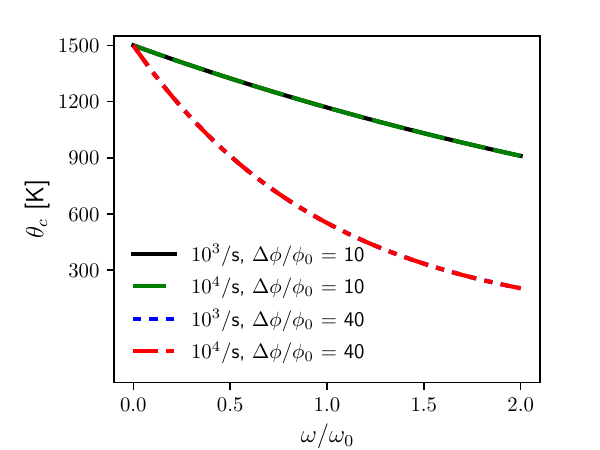}\label{fig5g}} \qquad
    \subfigure[band energy]{\includegraphics[width=0.32\textwidth]{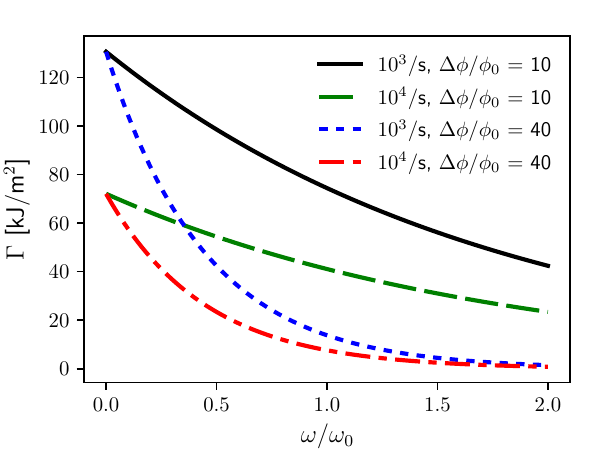}\label{fig5h}}
  \end{center}
  \vspace{-0.5cm}
\caption{Influence of degradation factor $\omega$ on shear-band characteristics, two defect concentrations $\Delta \phi / \phi_0$, $\omega_0 =  0.025$, other properties for HY-100 steel of Table~\ref{table1}:
(a) $\gamma_c$ 
(b) $\gamma_b$
(c) $a$
(d) $\delta_b$
(e) $\zeta_b$
(f) $t_c$
(g) $\theta_c$
(h) $\Gamma$
}
\label{fig5}       
\end{figure}

Outcomes for $\omega / \omega_0 \in [0,2]$ at fixed $\eta = \eta_0$ are reported in Fig.~\ref{fig5}.
All predicted quantities decrease with increasing defect-softening parameter $\omega$.
Rates of decrease are logically more pronounced when $\Delta \phi / \phi_0$ is increased, here from 10 to 40.
Critical strain in Fig.~\ref{fig5a} decreases modestly as strain rate increases from $10^3$ to $10^4$/s, whereas
the effect of $\dot{\gamma}$ on $\gamma_b$ is nearly negligible in Fig.~\ref{fig5b}.
Shear band width $a$, slipped distance $\delta_b$, process zone length $\zeta_b$, growth time $t_c$,
and surface energy $\Gamma$ in respective Figs.~\ref{fig5c}, \ref{fig5d}, \ref{fig5e}, \ref{fig5f}, and \ref{fig5h} all show similar trends with changes in $\omega$, $\dot{\gamma}$, and
$\Delta \phi / \phi_0$. Terminal band temperature in Fig.~\ref{fig5g} is again independent of $\dot{\gamma}$.%applied strain rate, but it does decrease with increased defect-softening.

\subsection{Discussion: model utility and limitations}

As shown in Section 5.2, the present framework with viscosity and defect susceptibility enables a closer
depiction of shear band width $a$, and other quantities to a limited extent, than prior analytical models.
This benefit suffers the cost of the additional parameter $\eta$ for nonlinear viscosity and the function $\lambda$
for effects of local defects, or more broadly, microstructure heterogeneity.
In the present application, $\eta$ and the constant $\omega$ entering $\lambda$ of \eqref{eq:lambdadef}
are obtained by calibration of critical strain $\gamma_c$ to numerical data \cite{vishnu2022}.
The shear band width $a$ in Tables~\ref{table2} and \ref{table3} is then a model prediction, rather than calibration.
However, if $\gamma_c$ is unknown or of no concern, and if $a$ is measured experimentally, 
parameters can be obtained using an alternative sequence.
From \eqref{eq:opt} and \eqref{eq:alphasat}, an implicit relation for viscosity $\eta$ and heterogeneity function $\lambda$ can be derived if $a$ is known:
\begin{equation}
\label{eq:visce}
\frac{[\lambda(\chi_0 + \eta \dot{\gamma}_0 / \rho_0)]^5}{(\chi_0 + 2 \eta \dot{\gamma}_0 / \rho_0)^2}= \frac{\beta^2 \tau_y^3 \dot{\gamma}}{9 c^2 \rho_0^3 f_M^2 \theta_M^2} a^4.
\end{equation}
If a material were perfectly homogeneous (possible in simulations \cite{vishnu2022}, but not real experiments), then 
$\lambda = 1$ and $\eta$ can be calculated from \eqref{eq:visce} by simple numerical iteration.

The greatest source of uncertainty, and thus most severe deficiency, of the present framework, is the 
function $\lambda$. The exponential form \eqref{eq:lambdadef}, or ranges of $\omega$, should not be arbitrarily 
generalized to all porous microstructures. Furthermore, the defect concentration $\Delta \phi / \phi_0$
defined in Appendix B is pragmatically restricted to small average porosities $\phi_0$ (e.g., $\phi < 1$ requires $\phi_0 \lesssim 1\%$
if $\Delta \phi / \phi_0 \leq 100$ as explored in Section 5.2). Experimental and/or numerical simulation data on shear band characteristics such as strain or time to failure
versus extreme-value statistics of structural or state features \cite{zhang2023} that influence localization (e.g., distributions of porosity \cite{batra1994,sun2011,marvi2021}, grain orientations \cite{molinari1988}, weakening inclusions \cite{karim2021}, and dynamically recrystallized grains \cite{rittel2008}) are required to inform a more universal, and more stringently validated, function $\lambda$ broadly relevant to viscoplastic polycrystals.  Like any isotropic continuum plasticity-damage approach \cite{gurson1977,batra1994,nahshon2008,kubair2015,wright2002}, effects of discrete defects and individual grains are unavoidably smeared into homogenized properties and statistical representations, even as void or grain sizes can affect critical strains and shear band widths \cite{vishnu2022}. Thermodynamics and $\lambda$ could be extended to explicitly capture mechanisms such as 
variable fraction of energy of cold work \cite{rittel2006b,kunda2024} and DRX 
\cite{rittel2008,longere2018,chen2025}, but some detail must likely be compromised to enable simple and useful closed-form expressions. %Extension to loading different from simple shear is also needed for wider relevance. Porosity should evolve more strongly in tension or compression \cite{silva1997}.

However, given $\lambda$ and its limitations, the present research does provide a tractable framework, motivated by fundamental mechanics principles and with closed-form analytical expressions, 
enabling quantification and exploration of effects of defects and other material properties (e.g., yield strength, density, thermal diffusivity, viscosity) on features of shear bands, extending the seminal works of Grady and Kipp \cite{grady1985,grady1987,grady1991,grady1992,grady1994}. Tradeoffs among variations in $\lambda$ (e.g., pores arising from contemporary AM techniques) and other physical properties can be optimized to mitigate shear banding by increasing time to collapse $t_c$ or shear-band resistance energy $\Gamma$, for example.

%\pagebreak
\section{Conclusions}
A model of shear band evolution extends the pioneering treatments of Grady and Kipp
to include nonlinear viscous effects and initial defects.
Non-Newtonian viscosity delays stress collapse and widens the band similarly
to heat conduction. Local defects accelerate collapse, narrow the band, and
reduce temperature rise. Simplifying idealizations allow
calibration to, and comparison with, dynamic torsion data from 
numerical and experimental studies on porous and fully dense metals. 
Upon introduction of only two calibrated parameters,
the presently derived formulae better match experimental trends for critical strain, local
band strain, local temperature, and shear band thickness than the original 
rate-independent formulae of Grady and 
formulae for shear band width derived in past analytical models. Unlike the current treatment, 
none of these prior formulae included
quantitative measures of defects such as porosity. 
This work establishes a basic framework within which effects of processing defects
 on shear-band failure in rate-sensitive materials can be analyzed.
The need for more
experimental and numerical data relating local defect distributions in the microstructure to 
the onset and physical characteristics of shear bands has been emphasized.
%\pagebreak
\appendix

\setcounter{figure}{0} \renewcommand{\thefigure}{A.\arabic{figure}}
\setcounter{table}{0} \renewcommand{\thetable}{A.\arabic{table}}
\setcounter{section}{0} \renewcommand{\thesection}{A.\arabic{section}}
\renewcommand{\theequation}{A.\arabic{equation}}
\setcounter{equation}{0}

\section*{Appendix A: Conditions at initial instability}

Denote shear stress, shear strain, shear strain rate, and temperature by
$\tau$, $\gamma$, $\dot{\gamma}$, and $\theta$. Consider a power-law flow rule 
for an isotropic rigid-viscoplastic solid \cite{fressengeas1987,molinari1987,claytonJMPS2024} ($\gamma$ := total $\approx$ plastic shear):
\begin{align}
\label{eq:flowstress}
 \tau(\gamma, \dot{\gamma},\theta;\phi_0)   = (1-\phi_0) g_0 
 \biggr{(} 1 + \frac{\gamma}{ \gamma_0} \biggr{)}^n  \biggr{(} \frac{\theta }{ \theta_0} \biggr{)}^\nu \biggr{(} \frac{\dot{\gamma}}{\dot{\gamma}_0} \biggr{)}^m.
\end{align}
Average porosity is $\phi_0$, assumed constant here.
Linear strength reduction by the factor $1-\phi_0$ is standard among ductile plasticity-damage models
when $\phi_0$ is small relative to unity \cite{gurson1977,becker1988,claytonNCM2011,claytonJDBM2021}.
Strain hardening exponent $n$ is usually positive, thermal softening exponent $\nu$ usually
negative. Rate sensitivity is $m > 0$, reference temperature is $\theta_0$, and $\gamma_0> 0$ and $\dot{\gamma}_0 > 0$ are constants.
Reference yield stress is the constant $g_0 > 0$.
Under simple shear deformation, an insulated homogeneous material has spatially constant $\gamma$ and $\theta$ fields and obeys the local energy and mass balances
\begin{equation}
\label{eq:ebalance}
(1-\phi_0) \rho_0 c = \beta \tau \dot{\gamma}, \qquad \rho = (1-\phi_0) \rho_0.
\end{equation}
Mass density of the incompressible matrix material without voids is $\rho_0 > 0$, specific heat per unit mass is $c > 0$,
and the Taylor-Quinney factor is $\beta \geq 0$, all simply constants.

Substituting \eqref{eq:flowstress} into \eqref{eq:ebalance}, eliminating the factor $1 -\phi_0$, and defining $c_V = \rho_0 c$ the specific heat per unit volume, 
\begin{equation}
\label{eq:dtheta}
\d \theta = \frac{ \beta g_0}{c_V} 
 \biggr{(} 1 + \frac{\gamma}{ \gamma_0} \biggr{)}^n  \biggr{(} \frac{\theta }{ \theta_0} \biggr{)}^\nu \biggr{(} \frac{\dot{\gamma}}{\dot{\gamma}_0} \biggr{)}^m \d \gamma.
\end{equation}
For constant $\dot{\gamma}$, \eqref{eq:dtheta} is integrated for $\theta = \theta(\gamma)$ by separation of variables \cite{molinari1987,claytonJMPS2024},
with $\theta_i = \theta(0)$:
\begin{align}
\label{eq:thetan}
%   {\theta}^{-\nu} \d {\theta}  = \frac{\beta}{c_V \theta_0^\nu} g_0 \omega \biggr{(} 1 + \frac{\gamma}{ \gamma_0} \biggr{)}^n  \biggr{(} \frac{\dot{\gamma}}{\dot{\gamma}_0} \biggr{)}^m  \d \gamma \quad \Rightarrow \quad 
%\\ 
 \theta (\gamma)  = 
{\pmb{ \biggr{[} }} \theta_i^{1-\nu} + \frac{(1-\nu)  \beta g_0  \gamma_0} {(1+n) c_V \theta_0^\nu} 
 \biggr{(}  \frac{\dot{\gamma}} {\dot{\gamma}_0} \biggr{)}^m
\biggr{\{} \biggr{(}1 + \frac{\gamma}{ \gamma_0} \biggr{)}^{1+n}-1 \biggr{\}} 
{\pmb{ \biggr{]}} }^{\textstyle{\frac{1}{1-\nu}}}.
\end{align}

Insertion of \eqref{eq:thetan} into \eqref{eq:flowstress} gives a stress-strain function $\tau = \tau(\gamma)$
at fixed $\dot{\gamma}$ and $\phi_0$.
Minimum strain at which localization initiates from a small perturbation is
identified with the peak instability strain ${\gamma}_p$ \cite{wright2002,yan2021,shawki1988,anand1987}.
In simple shear, $\gamma_p$ 
is $\gamma$ for which $ \d \tau / \d \gamma = 0$ (i.e., where $\tau (\gamma)$ attains a local maximum). 
When $\theta_i =  \theta_0$, the following implicit solution for instability strain ${\gamma}_p$ is derived \cite{claytonarx2025}:
\begin{align}
\label{eq:instrain}
&{\gamma}_{p} = \underset{\gamma \geq 0 }{\text{arg} \, 0} \left[ 
\frac{n}{\nu A_p}
+ \frac{(1 + \frac{ \gamma }{ \gamma_0})^{1+n}}{1 + A_p (\frac{1-\nu}{1+n}) \{ (1 + \frac{ \gamma }{ \gamma_0})^{1+n} - 1\} }
 \right], \qquad A_p = \frac{\beta g_0 \gamma_0 }{c_V \theta_0} \biggr{(} \frac{\dot{\gamma}}{ \dot{\gamma}_0} \biggr{)}^m.
 \end{align}
 
%\pagebreak
\appendix

\setcounter{figure}{0} \renewcommand{\thefigure}{B.\arabic{figure}}
\setcounter{table}{0} \renewcommand{\thetable}{B.\arabic{table}}
\setcounter{section}{0} \renewcommand{\thesection}{B.\arabic{section}}
\renewcommand{\theequation}{B.\arabic{equation}}
\setcounter{equation}{0}

\section*{Appendix B: Porosity data and defect concentrations}
Average porosity $\phi_0$, defect concentration $\Delta \phi / \phi_0$, and critical strain $\gamma_c$
(titanium and HY-100 steel) are obtained
from numerical data on ten realizations R1, $\ldots$, R10 of AM microstructure INC1Z of Ref.~\cite{vishnu2022}.
Values of $\phi_0$ are sourced directly from Table 3 of Ref.~\cite{vishnu2022}. Critical strains at localization
$\gamma_c$ for each realization and each material are obtained from Fig.~6(b) of Ref.~\cite{vishnu2022}.
Values are given in Table~\ref{tableB1}.
Results of Ref.~\cite{vishnu2022} show a decrease in $\gamma_c$ with increasing $\phi_0$ and
increasing $D_m$, where $D_m$ is maximum diameter of all spherical voids in a realization. However, $D_m$ and
$\phi_0$ are not uncorrelated: data suggest a general increase in $D_m$ with increasing $\phi_0$.
If sizes obey a random distribution, the probability of a larger void increases as their number per unit volume $n_v$ increases.

Less correlation with $\phi_0$ is attained when a normalized defect concentration  $\Delta \phi / \phi_0$ is used:
\begin{equation}
\label{eq:phiconc}
\frac{\Delta \phi}{\phi_0} = 
\begin{cases}
& ({\phi_m - \phi_0})/{\phi_0} =  \frac{\pi}{6} n_v D_m^3 / \phi_0   - 1, \quad [\phi_0 > 0, \Delta \phi \in (0,1-\phi_0)]
\\  & 0, \qquad \qquad \qquad \qquad \qquad \qquad \qquad \qquad \qquad [\Delta \phi = 0].
\end{cases}
\end{equation}
Limits in \eqref{eq:phiconc} restrict maximum local porosity to $\phi_m < 1$ and enforce a zero value when perturbations vanish, regardless of $\phi_0$. 
For the former, a maximum $\Delta \phi / \phi_0 \approx 100$ necessitates $\phi_0 \lesssim 1 \, \%$.
Values in Table~\ref{tableB1} are calculated using $\phi_0$, $D_m$ and $n_v$ from Table 3 of Ref.~\cite{vishnu2022}. Average porosity $\phi_0$ is not an appropriate influencing function for localization susceptibility in the current model. In the
limit that pores are infinitesimal in size and homogeneously distributed, $\tau_y$ should be uniformly reduced 
(e.g., by a factor of $1-\phi_0$), but no impetus exists for localization at any one location.
\begin{table}%[ht!]
\footnotesize
\caption{Porosity data and critical strain $\gamma_c$ at $\dot{\gamma} = 10^3$/s: numerical realizations R1, $\ldots$ , R10 of Ref.~\cite{vishnu2022}}
\label{tableB1}       % Give a unique label
\centering
\begin{tabular}{lcccccccccc}
\hline\noalign{\smallskip}
Quantity & R1 & R2 & R3 & R4 & R5 & R6 & R7 & R8 & R9 & R10  \\
\noalign{\smallskip}\hline\noalign{\smallskip}
$\phi_0$ [\%] & 0.075 & 0.039 & 0.059 & 0.073 & 0.046 & 0.114 & 0.06 & 0.043 & 0.046 & 0.132 \\
$\Delta \phi / \phi_0$ & 21.5 & 37.0 & 42.8 & 15.3 & 7.4 & 42.1 & 9.2 & 39.2 & 11.1 & 36.8 \\
$\gamma_c$ (Ti$^\dagger$) & 0.26 & 0.27 & 0.245 & 0.255 & 0.345 & 0.205 & 0.34 & 0.27 & 0.35 & 0.215 \\
$\gamma_c$ (HY-100) & 0.41 & 0.425 & 0.375 & 0.4 & 0.57 & 0.315 & 0.525 & 0.415 & 0.56 & 0.35 \\ 
\noalign{\smallskip}\hline 
$^\dagger$$\gamma_c = 0.55$ for $\Delta \phi = 0$
\end{tabular}
\end{table}

%\pagebreak
%\small
%\footnotesize
\bibliography{refs}
%
%\clearpage
%\pagestyle{empty}
%\normalsize
%\input{sreply}
%
\end{document}